\documentclass[12pt, a4]{elsart}
\usepackage{graphicx,natbib}
\usepackage{amsmath}
\usepackage{amsfonts}
\usepackage{amsbsy}
\usepackage{amssymb}

\journal{Icarus}

\newcommand{\x}{\ensuremath{\mathbf{x}}}
\renewcommand{\v}{\ensuremath{\mathbf{v}}}
\renewcommand{\u}{\ensuremath{\mathbf{u}}}
\renewcommand{\d}{\ensuremath{\partial}}
\newcommand{\w}{\ensuremath{\mathbf{w}}}

\renewcommand{\ln}{\ensuremath{\text{ln}\,}}

\newcommand{\eee}{\ensuremath{\mathbf{e}}}
\newcommand{\ex}{\ensuremath{\mathbf{e}_{x}}}
\newcommand{\ey}{\ensuremath{\mathbf{e}_{y}}}
\newcommand{\ez}{\ensuremath{\mathbf{e}_{z}}}

\begin{document}

\begin{frontmatter}

\title{The linear stability of dilute particulate rings}
\author[cam1]{Henrik N. Latter\corauthref{cor}},
\corauth[cor]{Corresponding author.}
\ead{hl278@cam.ac.uk}
\author[cam1]{Gordon I. Ogilvie}
\ead{gio10@cam.ac.uk}

\address[cam1]{DAMTP, University of Cambridge, CMS, Wilberforce Road, Cambridge CB3 0WA, UK}

\begin{abstract}
Irregular structure in planetary rings
 is often attributed to the intrinsic instabilities 
of a homogeneous state undergoing Keplerian shear. Previously these have been analysed 
with simple hydrodynamic models. We instead employ a kinetic theory, in which
we solve the linearised moment equations derived in Shu and Stewart
1985 for a dilute ring. This facilitates an examination of
 velocity anisotropy and
non-Newtonian stress, and their effects on the viscous and viscous/gravitational
 instabilities thought to occur in
 Saturn's rings. Because we adopt a dilute gas model, the
 applicability of our results to the actual dense rings of Saturn
 are significantly curtailled. Nevertheless this study is a
 necessary preliminary before an attack on the difficult problem of
 dense ring dynamics.

We find the Shu and Stewart formalism admits analytic stability 
criteria for the viscous overstability, viscous 
instability, and thermal instability. These criteria are compared with those of a 
hydrodynamic model incorporating the effective viscosity and cooling 
function computed from the kinetic steady state. We find the 
two agree in the `hydrodynamic limit' (i.e. many collisions 
per orbit) but disagree when collisions are less frequent, when we expect the
viscous stress to be increasingly non-Newtonian and 
the velocity distribution increasingly anisotropic. In particular, hydrodynamics predicts 
viscous overstability for a larger portion of parameter 
space. We also numerically solve the linearised 
equations of the more accurate Goldreich and Tremaine 1978
kinetic model and 
discover its linear stability to be qualitatively the same 
as that of Shu and Stewart's. Thus the simple collision 
operator adopted in the latter would appear to be an adequate 
approximation for dilute rings, at least in the linear regime.

\end{abstract}

\end{frontmatter}

\section{Introduction}

This paper examines the collisional dynamics of
differentially rotating, particulate disks, and addresses specifically
the intrinsic mechanisms
which might induce the generation of irregular, fine-scale
structure, such as that exhibited by Saturn's B-ring.  \emph{Voyager}, and \emph{Cassini} more recently, revealed that these
essentially axisymmetric
variations appear on a very wide range of length-scale-- from the limit of
resolution ($\sim$ 100m) to several hundred kilometres 
(Horn and Cuzzi 1995, Porco \emph{et al.} 2005 ). Though some
proportion of B-ring structure may be caused by non-dynamical effects such as variations in albedo and phase
function- see Cuzzi and Estrada 1996, for example) and from gravitational interactions with neighbouring
satellites (Thiessenhusen \emph{et al.} 1995), it is believed that the majority 
corresponds to changes in optical thickness resulting from a collective
dynamics (Tremaine 2003).

This work studies \emph{dilute} particulate systems, and as
 such may not be directly applicable to the saturnian rings.
 It is nevertheless an informative first stop before we embark on the challenging stability
analysis of a dense granular gas. Other than presenting a general
framework in which such a calculation can be tackled, the dilute
gas formalism 
permits us to isolate analytically the interesting effects of
anisotropy and non-Newtonian viscous stress upon the local axisymmetric instabilities
 thought to be manifest in planetary rings. 

Since \emph{Voyager} first reported the radial stratification of
 Saturn's rings, theoreticians
 have
 advocated a number of possible causes. 
These have included: ballistic transport (Durisen 1995),  electromagnetic effects (Goertz and Morfill 1988), and
 interleaved shearing and shear-free zones (Tremaine 2003). 
Others have looked to the
 local instabilities of viscous fluid disks, interpreting the
 disordered state observed as the saturated
endpoint of their nonlinear evolution. The
`viscous instability' was the first proposed (Lin and Bodenheimer
1981, Ward 1981, and Lukkari 1981).
Essentially a monotonic `clumping', it is associated with an outward angular
momentum flux which decreases with surface density: $d(\nu \sigma)/d\sigma
<0$ (where $\nu$ is kinematic viscosity and $\sigma$ surface mass density).
Though a dilute ring's viscosity depends on optical depth in a manner
which promises the existence of such an instability
(Goldreich and Tremaine 1978, Shu
and Stewart 1985), Saturn's rings are
most likely `dense', and theoretical and numerical N-body studies have revealed that
such rings do not manifest the appropriate viscosity for this
instability to develop (Araki and Tremaine 1986, and Wisdom and
Tremaine 1988).

The `viscous overstability' was first examined in the context of accretion
disks by Kato (1978), and, as the name suggests, originates in
overcompensation by the system's restoring forces: the
stress oscillation which accompanies the epicyclic response
in an acoustic-inertial wave will force the system back to
equilibrium so strongly that it will `overshoot'. The mechanism relies
on: a) the
synchronisation of the viscous stress's oscillations with those of
density, and b) the viscous stress increasing sufficiently
in the compressed phase. In hydrodynamics only the latter
consideration is relevant, which furnishes the
criterion for overstability: $\beta\equiv
(d\ln\nu/d\ln \sigma)>\beta^*$, where $\beta^*$ is a number dependent
on the thermal properties of the ring (Schmit and Tscharnuter 1995). 

The viscous overstability has been
a favoured explanation for smaller scale B-ring structure in recent years,
a status stemming, primarily, from the
viscosity profiles computed by Araki and Tremaine's 1986 dense gas model and Wisdom
and Tremaine's 1988 particle simulations. Both appear to satisfy the
above criterion. Consequently the linear behaviour of the instability has been thoroughly
examined, though only within a hydrodynamic framework (Schmit and
Tscharnuter 1995, Spahn \emph{et. al.} 2000, and Schmidt \emph{et al.} 2001). In addition, Schmidt and Salo (2003)
have constructed a weakly nonlinear theory, and the overstability's long-term,
nonlinear behaviour has been numerically studied by Schmit and
Tscharnuter (1999). An isothermal model was adopted in both cases.
\vskip0.3cm

However, fluid mechanics is only one of several theoretical approaches
thathave been deployed to capture the
behaviour of planetary rings: N-body
simulations, generalisations of
stellar dynamics, and kinetic theory. This plurality is surely indicative of how the field of planetary
ring dynamics falls uncomfortably between the more familiar frameworks of
classical physics. Our analysis will predominantly focus upon the
kinetic approach, and solve
the set of moment equations derived from the kinetic theories proposed
 in Shu and Stewart 1985 (hereafter referred to as `SS85' ) and in Goldreich
and Tremaine 1978 (referred to as `GT78'). 
These formalisms have hitherto been used mainly to
establish equilibrium solutions. This paper goes the next step by
exploring their linear theory. 

Such an analysis is more involved
than the analogous hydrodynamic calculation; however, the latter's
adoption of the Navier-Stokes stress model introduces two assumptions
which may be inappropriate in the ring context
and whose consequences are instructive to investigate. Firstly, the
Navier-Stokes model presumes the distribution of the particles' velocity
dispersion to be nearly isotropic. In the regime of many collisions per orbit
this is an acceptable supposition, as collisions scatter particles randomly
on the average. However if the collision rate, $\omega_c$, is of the same
order as the orbital frequency, $\Omega$, (as it is presumed to be in Saturn's
rings) this need not be true. Secondly,
hydrodynamics assumes an `instantaneous' (local in time) relationship
between stress and strain. This may not hold
 when
$\omega_c\sim\Omega$. Generally the viscous stress possesses
 a relaxation time of order $1/\omega_c$, which in this regime will be comparable
to the dynamic time scale. Thus the immediate history of the
stress cannot be ignored.
Its inclusion should have the most impact on the stability of oscillating
modes, especially the overstability, it depending on the synchronisation of
the stress and density oscillations. A kinetic model can address both
issues, accounting for anisotropy within an appropriate collision term and
providing a straightforward way, by the taking of moments, to generate
dynamical equations for the viscous stress. Another advantage is that
a kinetic model lets us explicitly include the microphysics of
particle-particle interactions and thence to potentially model a larger set of the
physical mechanisms at play (such as collisions, irregular surfaces,
spin , size distribution, etc). It also narrows the scope of our simplifying
assumptions to the particulars of collisions between spheres of ice, which
have been observed in the laboratory (see Bridges \emph{et al.} 1984,
Hatzes \emph{et al.} 1988, amongst others) . 
\vskip0.3cm
The paper is organised as follows. In Section 2 we present a brief
derivation of the set of closed, vertically averaged moment equations in a local
model, the shearing sheet, for a general kinetic equation. We then
specialise to the SS85 kinetic model and briefly exhibit its
equilibrium characteristics. Section 3 contains the linear theory of a
general kinetic equation and proves results pertaining to self-gravity 
and the thermal modes when in the limit of length-scales much longer than the disk
height. 
We then derive stability criteria
for the viscous instability and overstability specific to the SS85 model.
Comparisons with an analogous calculation for a hydrodynamic
model and the GT78 model are performed in Sections 4 and
5. In the former we isolate the effects of non-Newtonian
stress on the linear theory, and in the latter observe how the SS85 collision term
approximates satisfactorily that of the more accurate triaxial Gaussian model. We
draw our conclusions in Section 6.

\section{Kinetic Model}

\subsection{Simplifying Assumptions}
The formulation of a suitable kinetic theory poses a number
of difficulties, namely the closure of the moment hierarchy and the
simplification of the collision term. But the various approximations
these require by no means cripple the approach.
The most fundamental assumption we make is that our ring is
composed exclusively of hard, identical, and indestructible spheres. In
addition we presume that the particles are non-spinning. These
simplifications render the mathematics convenient (but see Stewart
\emph{et al.} 1984 and Shu and Stewart 1985 for justifications)-- now the
inelasticity of the collisions is quantified solely by the normal
coefficient of restitution, $\varepsilon$, which is the ratio of the
normal relative speed after and before a collision. It is generally a function
of normal impact velocity, $v_{n}$, and possibly other parameters like ambient
temperature and
pressure, particle size and mass (Hatzes \emph{et al.} 1988,
Dilley 1993) .

 A number of laboratory experiments have
been conducted in order to ascertain the collisional properties of ice
spheres in saturnian conditions, in particular the relationship
between the coefficient of restitution and $v_n$ (see Bridges \emph{et
  al.} 1984, Hatzes \emph{et al.} 1988, Supulver \emph{et al.} 1995, and
Dilley and Crawford 1996). Excepting Dilley,
these authors successfully fit a step-wise power law to their data for collisions sufficiently gentle and/or surfaces
sufficiently frosted: 
\begin{align} \label{bridge}
\varepsilon(v_{n}) = \begin{cases} (v_{n}/v_c)^{-p}, 
&\text{for}\quad v_n > v_c, \\
 1, & \text{for} \quad v_n \leq v_c,
\end{cases}
\end{align}
where $v_c$ and $p$ are parameters contingent on the material properties of the
ice balls and their ambient environment.
Bridges' data admit $p=0.234$ and
$v_c=0.0077$ cm s$^{-1}$ for frosted
particles of radius 2.5cm at a temperature of 210K (significantly higher
than the appropriate conditions). Hatzes finds $p=0.20$ and $v_c=0.025$
cm s$^{-1}$ for the case of frosted particles at 123K (however, for the
case of smoother particles an exponential law
provides a better fit). At slightly lower
temperatures ($\approx$100K)
 Supulver \emph{et al.} obtain $p=0.19$ and $v_c= 0.029$
 cm s$^{-1}$ with their torsional pendulum fixed.
Though the functional relationship
\eqref{bridge} is useful, 
 the coefficient of restitution varies considerably as the physical
 condition of the contact surface is more or less frosty or
 sublimated,
 as Hatzes \emph{et al.} explore. Also the neglect of spin, the
 effects of glancing collisions,
 irregularly shaped surfaces, coagulation and erosion, mass transfer and the role of amorphous ice,
further emphasises the simplicity of the collision model adopted.
 Studies
of those processes we omit are found in
 Salo 1987, Araki 1991 and Morishima and Salo 2006 (theory and simulations), and in McDonald \emph{et al.} 1989, Hatzes
\emph{et al.} 1991, Supulver \emph{et al.} 1995 and Supulver
\emph{et al.} 1997 (experiments).

From this point on we denote by $\varepsilon$ the
\emph{averaged} coefficient of restitution, which is a function
of the macroscopic variable, $c$, the velocity dispersion. For a
general step-wise power law
 and a Maxwellian velocity distribution,
 it is straightforward to obtain an analytic expression for $\varepsilon$ averaged over
   collisions. Its functional dependence on $c$ is,
 unsurprisingly, the `smoothed' analogue of
 Eq.\eqref{bridge}'s piecewise dependence on $v_n$. 
  
The rings we study are \emph{dilute}. 
We realise this curtails the
applicability of our results considerably as Saturn's rings, even the
optically thin regions, are thought to exhibit important dense
effects. 
Araki and
Tremaine (1986) revealed significant
collisional contributions to the kinetic equilibrium
when collisions are sufficiently inelastic;
 and the experimentally derived elasticity laws of
\eqref{bridge} predict collisions are dissipative enough for this
`cold' regime to be widespread in Saturn's rings, given appropriate
particle sizes (see Marouf \emph{et al.} 1983 and Zebker
\emph{et al} 1985). This conclusion has been confirmed by particle
simulations, Wisdom and Tremaine 1988 and Salo 1991 the most notable.
Moreover, both simulations and theoretical models show that the
dynamics of small particles (for which nonlocal effects are less
important) strongly couple to the dynamics of the largest
particles (for which nonlocal effects \emph{are} important), both in
two size systems and in
polydisperse disks exhibiting power-law size
distributions akin to Saturn's (see Stewart \emph{et al.} 1984, Salo 1987, Lukkari 1989, Salo
1991, and Salo 1992b). 

Physically then, a dilute model best describes planetary rings in which the
radii of the largest particles are of the order of 
centimetres, or tens of centimetres. Estimates on the maximum
size of particle radius are easy to establish from the dilute
theory, though these are only \emph{necessary} bounds: sufficient conditions for the applicability of the dilute
model can only be determined from the dense theory. At lower
optical depths, however, we expect the dilute estimates to serve as a
reasonable guide.
   We obtain these by combining
 the famous equilibrium relation between optical thickness, $\tau$, and
 $\varepsilon$, computed in GT78 (cf. Fig 7), with the
 elasticity laws embodied in \eqref{bridge}. This determines the
 dependence of velocity dispersion
  on $\tau$. We then suppose that non-local
 effects may be neglected if a particle's r.m.s. speed, $\sqrt{3}c$, is an order larger
 than the shear velocity across two particle radii, $3\Omega a$, where
 $a$ is particle radius (Araki 1991)-- the ratio of these quantities is a convenient
 measure of `nonlocality', but only for optical thicknesses less than
 about 1 or 2 (see Salo 1991). Thereupon
 we compute an
 upper limit on $a$ below which the dilute model is acceptable. If $\tau=0.2$ at
 C-ring distances, $a_{\text{max}}$ ranges from 31cm (for Supulver
 \emph{et al.}'s data) to 5.3cm (for Bridges \emph{et al.}'s data).
 At A-ring distances for $\tau=0.5$ the maximum particle radius is
 33cm 
(Supulver \emph{et al.}) or 6.3cm (Bridges \emph{et al.}).
 At B-ring distances and $\tau=1$, the Bridges \emph{et al.}
 data imply $a_{\text{max}}\approx 2.2$cm, or, alternatively,
 $v_c\approx 0.35$ cm s$^{-1}$ for
 metre-sized particles. This last estimate appears consistent with those
 simulations of Salo (1991) in which $v_c$ was increased to 20 and
 40 times $v_B=0.01$ cm/s (cf. his Fig.'s 5a, 5c and 5d). It is interesting to note
 the sensitivity of the critical $a$ to the elasticity law
 adopted: relations which predict slightly less dissipative impacts
 allow a substantially higher velocity dispersion, and subsequently
 $a_{\text{max}}$.      

Our dilute ring is self-gravitating. Disk self-gravity may be decomposed into three effects.
Firstly, inter-particle gravitational forces will lead to an additional source of
(elastic) scattering via gravitational encounters. They will also enhance the collision
frequency of physical collisions (gravitational focusing).
 For either process to be appreciable, the velocity dispersion
must be of
order the particles' escape velocity, i.e.
$G\rho a^2 \sim c^2$, where $G$ is the gravitation constant and $\rho$ the
mass density of the particle material.
 This can be recast as $(a\Omega)/c \sim 1$, for typical values of the
saturnian system. As non-local effects are neglected, the left side of
this scaling will be $\ll 1$, and hence gravitational scattering and focusing may be omitted. Secondly, the aggregate of
all the particle attractions will increase the disk's
vertical restoring force and further flatten the disk. Subsequently the
collision frequency will increase, an effect which we
include. Thirdly, there will arise a dynamical contribution to the disk's
stability which issues from the extra term in the momentum equation. Though this
term will be unimportant at equilibrium, being dominated by the
planet's gravitational potential, it will be significant in the linear
theory. As is well known, an inviscid fluid disk collapses under the
action of self-gravity for a Toomre parameter $Q$ less than 1, its
 unstable modes growing on a band of
intermediate wavelengths, with rotation and pressure stabilising the long and
short scales respectively (Toomre 1964, Julian and Toomre 1966). A
viscous disk however paints a
slightly more complex picture. Then it is better to think of self-gravity as `extending' the
viscous instabilities we discuss into larger areas of parameter space.
In these `extensions' the instabilities grow only in a certain
confined range of intermediate wavelength, unlike the non-self-gravitating case in which the longest wavelengths are the
first to become viscously unstable and overstable (Schmit and
Tscharnuter 1995). We shall examine this effect in some detail particularly
for the viscous instability. We should mention also that self gravity induces
transient, non-axisymmetric wakes, the analysis of which we omit (but see Salo 1992a).

\subsection{The Governing Moment Equations}
Consider a dilute gas of identical, smooth, inelastic spheres of
mass $m$ with phase space distribution $f(\x,\v, t)$. The number
of particles located in the volume $d\x$ centred at $\x$ with
velocities in the range $d\v$ centred at $\v$ at time $t$ is defined as
$  f(\x,\v,t)d\x\, d\v$. 
Taking moments of $f$ allows us to calculate the familiar macroscopic
characteristics of the gas.
Number density, $n$, bulk velocity, $\u$, and velocity dispersion, $c$, are defined through
\begin{align}
n(\x,t) &\equiv \int  f\, d \v, \\
n \u(\x,t) &\equiv \int  \v f\, d \v, \\
\frac{3}{2} n c^2(\x,t) &\equiv \int \frac{1}{2} |\v-\u|^2\, f d \v.
\end{align}   
 The phase space distribution 
satisfies a kinetic equation which is distinguished by a collision
operator, $(\d f/\d t)_c$, whose precise form reflects the collisional
microphysics. 

We
situate our gas in a \emph{shearing sheet}. This is a convenient 
representation of a
differentially rotating disk in which a small patch, centred on
a point moving on a circular orbit at
$r=r_{0}$, is represented as
a sheet in uniform rotation, $\Omega(r_0)\, \ez$, and subjected to a
linear shear flow, $\u_{0}=-2A_{0}x\, \ey$. The local rectilinear
coordinates $x$ and $y$ point in the radial and azimuthal directions
respectively, and
$A_{0}=-\frac{1}{2}r_{0}(d\Omega/d r)_0$; see Fig. 1.
For the dynamical analysis we consider the sheet
horizontally unbounded, at least compared to the vertical length scale
of the disk. The model hence provides an excellent approximation for rings
whose thickness is very small, such as Saturn's.

\begin{figure}
\begin{center}
\scalebox{.55}{\includegraphics{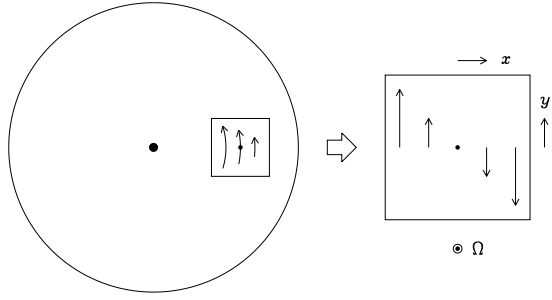}}
\caption{\footnotesize{The shearing sheet. The differential rotation
is locally represented as a rotation superimposed upon a linear shear flow.}}
\end{center}
\end{figure}

In terms of the peculiar velocity, $\w=\v-\u$, the kinetic equation
for such a patch of ring is
\begin{equation} \label{E3}
\frac{\d f}{\d t} + (w_{i}+u_{i}) \frac{\d f}{\d x_{i}} 
-\left[ \frac{\d u_{i}}{\d t} + (w_{j}+u_{j})\frac{\d u_{i}}{\d
    x_{j}} - F_{i} \right] \frac{ \d f}{\d w_{i}} = \left( \frac{ \d f}{\d t} \right)_{c},
\end{equation} 
where the force/unit mass is 
$$ F_{i}= -\frac{ \d (\phi_P+\phi_D)}{ \d x_{i}} - 2 \epsilon_{ijk} \Omega_{j}
  (w_{k}+u_{k}). $$
The appropriate centrifugal-gravitational potential of
  the planet is denoted by $\phi_P$, and
  the disk's gravitational potential by $\phi_D$. The tensor $\epsilon_{ijk}$ 
is the alternating tensor and the
  angular velocity is $\mathbf{\Omega}=
  \Omega_{0} \, \ez$, where $\Omega_0=\Omega(r_0)$. 
By multiplying \eqref{E3} by $1$, $ w_{i}$ and $
  w_{i}w_{j}$ and then integrating over all $\w$ we derive the continuity equation,
\begin{equation} \label{E4}
\d_{t} n + \d_{k} (n u_{k}) = 0,
\end{equation}
the equation of motion,   
\begin{equation} \label{E5}
n\left( \d_{t} u_{i} + u_{k} \d_{k} u_{i} \right) = -2n
\epsilon_{ijk} \Omega_{j} u_{k} - n \d_{i} (\phi_P+\phi_D) - \d_{j}p_{ij},
\end{equation}
and the pressure tensor equation,
\begin{align} \label{E6}
\d_{t} p_{ij} + u_{k}\d_{k}p_{ij} = -p_{ik}\d_{k}u_{j}
&-p_{jk}\d_{k}u_{i} -p_{ij}\d_{k}u_{k}
-2\epsilon_{ikl}\Omega_{k}p_{lj} \notag \\ & \hskip2cm -2\epsilon_{jkl}\Omega_{k}p_{li}  
-\d_{k} p_{ijk} + q_{ij}, 
\end{align}
where $p_{ij}$ and $p_{ijk}$ are the second and third-order moments,
$$ p_{ij\dots} = \int  w_{i} w_{j}  \dots f d \mathbf{w}, $$
and the collisional change in second moment is
$$q_{ij}= \int w_i w_j \left(\frac{\d f}{\d t}\right)_c d \mathbf{w}, $$
which is a yet to be specified function of the other field variables.
For notational brevity we have set $ \d_{i} = \d/\d x_{i}$.

\vskip0.3cm
As we are interested in scales $\gg H$ there is little point in
solving for a disk's detailed vertical structure. We hence vertically
integrate the moment equations, 
 a process which can be performed without much grief if it is assumed that
 $u_{x}$ and $u_{y}$ are effectively independent of $z$. 
The third order moments are neglected. This move closes the
system but restricts the
applicability of the equations to behaviour
on lengthscales much larger than the disk thickness. If the system exhibits variation on scales
close to $H$ the vertically integrated equations
will most likely predict incorrect behaviour.

We now have:
\begin{align} \label{E7}
D_{t} N  &= -N \d_{\alpha} u_{\alpha}, \\
N D_{t} u_{\alpha} &= -N
\d_{\alpha}(\Phi_P+\Phi_D) - 2N\epsilon_{\alpha z \gamma} \Omega_0
u_{\gamma} - \d_{\beta}P_{\alpha \beta}, \label{E8}\\
D_{t}P_{ij} &=  -P_{ij} \d_{\gamma}u_{\gamma} - P_{j\gamma}
\d_{\gamma}u_{i}
-P_{i\gamma}\d_{\gamma}u_{j}-2\epsilon_{izk}\Omega_0 P_{kj}\notag \\
& \hskip6.5cm  -2\epsilon_{jzk}\Omega_0 P_{ki} + Q_{ij}, \label{E10}
\end{align}
where $D_{t}= \d_{t}+ u_{\gamma}\d_{\gamma}$, Greek indices run
 only from $x$ to $y$ and upper case denotes vertical
 integration.
 Optical
 thickness is subsequently defined as $\tau=\pi\,a^2 N$.
To
 complete the set, expressions for $\Phi_P$ and $\Phi_D$ are required. We
 approximate the central body as perfectly spherical, which accounts
 for the former potential; the latter must be obtained from Poisson's
 equation:
\begin{equation} \label{Poisson}
\nabla^2 \phi_D = 4 \pi m G n,
\end{equation}
in which we may approximate $n$ by $N\delta(z)$, if it is assumed the disk is
 very thin.
 An additional equation for the mean
vertical displacement of the disk, $Z$, can be supplied from the $z$
 component of the equation of motion (see SS85), and the instances of $u_{z}$
in \eqref{E10} are identified with its rate of change,
$D_{t}Z$. But we shall not be investigating the vertical
warping of the disk and shall assume symmetry
about the plane $z=0$. Hence $Z=P_{xz}=P_{yz}=0$.

It is often illuminating to work with the internal energy and
 viscous stress equations in place of \eqref{E10}.  The former
 equation may be procured
 by taking half the trace of \eqref{E10} and defining $3 N C^2 \equiv P_{ii}$, where
 $C$ is the vertically averaged velocity dispersion.
A viscous stress equation proceeds by subtracting \eqref{E10} from
 $2\delta_{ij}/3$ times the energy equation,
if we define the viscous stress tensor as
$$ \Pi_{ij} \equiv N C^2 \delta_{ij} - P_{ij}. $$ 
We omit these details (see SS85).

\subsection{The SS85 Model}

We specialise to the modified
 BGK collision term (Bhatnagar \emph{et al.} 1954) which Shu and Stewart introduce in SS85:
$$ \left( \frac{\d f}{ \d t} \right)_{c} = \omega_{c} (f_{I}-f). $$ 
Here $\omega_{c}$ designates the collision frequency and $f_{I}$ is the
equilibrium distribution function to which the
collisional process tends to push the system. This is set 
equal to a Maxwellian. 
 Essentially the
model ensures that all particles which undergo
a collision at a certain time will be distributed immediately
afterwards like $f_I$. Given that on the average collisions randomize velocity
dispersion, this is not a bad approximation, but it does downplay the
particles' distribution immediately prior to the collision: if $f$ is
very far from a Maxwellian, the BGK model 
could overestimate the isotropizing power of collisions.  

In order to account for the
inelasticity of the particle collisions Shu and Stewart attribute to $f_{I}$
the same $n$ and $\u$ as $f$ (in order to conserve mass and momentum),
 but set its velocity dispersion $c_{I}^2$ less than $c^2$.
 They then assume
energy equipartition which permits them to write
\begin{equation} \label{inelastic}
 c_I^2=(2+\varepsilon^2)c^2/3,
\end{equation}
where the coefficient of restitution $\varepsilon$ is here understood to be
a quantity averaged over collisions. The approximation of energy
equipartition will worsen the further $f$ departs from Maxwellian, and
may explain the discrepancy between the results of SS85 and GT78 in
the collisionless limit.

To complete the model Shu and Stewart present an expression 
for the vertically averaged collision frequency. They find that, for a dilute, 
vertically isothermal disk
 with a locally Maxwellian velocity distribution,
\begin{equation} \label{colsig}
 \omega_c = 8 \Omega_z a^2 N \quad (= 8\Omega_z\tau/\pi)
\end{equation}
where $\Omega_z$ is the vertical epicyclic frequency. In this case the collision
frequency depends solely and linearly on $N$, and not on $C$. 
As it stands this expressions fails to capture the influence of
anisotropy. But it can be
improved on this count, somewhat, via inclusion of a factor $\sqrt{N
  C^2/P_{zz}}$, which comes from modifying the scale height of the disk
from $c/\Omega_z$ to $c_z/\Omega_z$, where $c_z$ is the vertical
velocity dispersion.

The influence of self-gravity on the vertical epicyclic
oscillation may be approximated through
$$ \Omega_z^2=\Omega^2_0 +(\d_z^2\phi_D)_{z=0}.$$
If we assume $\phi_D$ varies slowly with $x$ and $y$, then
$\nabla^2\approx\d_z^2$ in Poisson's equation, and with
 $n\propto \text{exp}(-z^2/(2H^2))$ 
this renders $\omega_c$'s dependence on surface number density
like:
\begin{equation} \label{ommie}
 \omega_c\approx 8a^2\,\Omega_0\,N \sqrt{1+ (2\sqrt{2\pi} Gm/H\Omega_0^2)\,N}.
\end{equation}
Hence $\omega_c$ can vary as steeply as $N^{3/2}$ in a
self-gravitating disk, and at equilibrium self-gravity enhances the
collision frequency by a factor $\sqrt{1+2\sqrt{2/\pi}\,Q^{-1}}$, where $Q$ is the
Toomre parameter.

\subsection{The Shu \& Stewart Equilibrium Disk}

The equilibrium homogeneous state of Keplerian shear can be computed analytically
 if the 
SS85 term is included, for which 
$$ Q_{ij} = \omega_c (N\,C_I^2 \delta_{ij} - P_{ij}).$$
 Let $\omega_{c}=\omega_{c}(N)$. Then, given a uniform density $N_{0}$ and
shear rate $\u_{0}= -\frac{3}{2} \Omega_0 \,x\, \ey$, the uniform viscous stress tensor 
 and the equilibrium value of $\varepsilon$ proceed from \eqref{E10}. 
The averaged
coefficent of restitution $\varepsilon$ depends on $C$ uniquely and, once its
equilibrium value is set, $C_{0}$ can be deduced from the average of Eq.\eqref{bridge}.
 The solution is
\begin{align} \label{eqm}
\mathbf{\Pi}_0 &= \frac{3 N_0 C^2_{0}}{11+2(\omega_0/\Omega_0)^2}
\left(\begin{matrix} -3 && -\omega_0/\Omega_0 && 0 \\
                              -\omega_0/\Omega_0        &&  2
                              && 0 \\
                              0 && 0 && 1 
                 \end{matrix}\right)
\end{align}
and 
\begin{equation}\label{E13}
\varepsilon_0 = \sqrt{ 1-
  \frac{9}{11+2(\omega_{0}/\Omega_0)^2}},
\end{equation}
for $\omega_{0}\equiv\omega_{c}(N_{0})$ and $\varepsilon_{0}\equiv\varepsilon(C_{0}^2)$. 

The form of $\Pi_{xy}$ demonstrates a local relationship
between shear and stress. Therefore one may write down the
effective kinematic viscosity,
\begin{equation}\label{E14}
\nu = \left(\frac{C_0^2}{\Omega_0}\right) \frac{ 2
(\omega_{0}/\Omega_0)}{11+2(\omega_{0}/\Omega_0)^2},
\end{equation}
as a function of $\omega_{0}$. Note the implicit dependence on
 $\omega_0$ through $C_0$: the equilibrium velocity dispersion depends
 on $\varepsilon_0$ which in turn is set by $\omega_0$. In the
 hydrodynamic limit,
 $\omega_{0}/\Omega_0\to \infty$, the viscosity must keep
 momentum flux constant and thus goes to zero like
 $(\omega_0/\Omega_0)^{-1}$ (Chapman and Cowling 1970). 
But in the collisionless limit the viscosity approaches $0$ like
 $\omega_{0}/\Omega$.
 In this regime the
 trajectories of particles between collisions are
long and subject to epicyclic confinement. As a consequence angular momentum
 transport is inefficient.
 Because the effective
 viscosity vanishes in both limits there exists a turning point
in the intermediate range when the gas is hot yet not too rarefied.
  A simple calculation shows that this turning point occurs at
 $\omega_0=\widetilde{\omega}$, where 
\begin{equation}\label{turn}
 \widetilde{\omega} = \tfrac{1}{2}\,\Omega_0\,\sqrt{ 9(1+2/e) + \sqrt{81(1+2/e)^2 +88}},
\end{equation}
the definition of which introduces the important parameter  
\begin{equation} \label{e}
e\equiv  (d\,\ln \varepsilon^2/d\, \ln C^2)_0.
\end{equation}
For a piecewise power-law, this parameter depends on both $p$ and $C_0/v_c$,
but only weakly for the latter: we can roughly approximate $e$ with $-p$.
Hydrodynamics would
      suggest that one look to the region
      $\omega_0>\widetilde{\omega}$  to observe viscous instability
      and to $\omega_0<\widetilde{\omega}$ for viscous
      overstability.
Because the system is dilute and the particles non-spinning there is
no effective bulk viscosity.

As Goldreich and Tremaine argue, energetic instability is assured if $e\geq 0$. Consider a small deviation in $C^2$
 \emph{below} its equilibrium value. When $e>0$ this change will generate a small decrease
 in $\varepsilon$ and a converse increase in collisional
 dissipation per collision (on the average). Thence cooling will slightly dominate viscous
 heating and thus
 amplify the disturbance, initiating a runaway loss of energy. A small
 deviation \emph{above} equilibrium 
 on the other hand
 causes a decrease in collisional cooling, which will precipitate a runaway
 heating. If $e=0$ the velocity dispersion cannot adjust either way
to balance the energy budget, and so the system will also be
 unstable. For $e<0$ the velocity dispersion will act to return the
 system to equilibrium if thermally disturbed, and the speed of
 return is
 proportional to the magnitude of
$e$. Isothermality in our model hence corresponds to $e\to -\infty$,
 as then
 any thermal perturbation will be
`instantly' quenched. Of course `instant quenching' is an artefact of
 the averaging process; in reality a particulate disk may return to
 thermal equilibrium no faster than the collision time, nor could an
 averaged $\varepsilon$ exhibit a steep enough dependence on $c$ to
 yield $e$ very large and negative.

  A more general perturbation including density
 displacements complicates the simple heuristic picture sketched above, but on
 long lengthscales the above argument should hold.

\section{Linear Stability Calculation}

\subsection{General Results}
In this section we 
establish the linear stability of the moment equations.
 We begin by presenting a number of results which hold for a
second-order system derived from a kinetic equation with an unspecified collision term. 

Consider the vertically integrated equations:
\begin{align} \label{p1}
D_t N &= -N \d_\alpha u_\alpha, \\
D_t u_\alpha &= -\d_\alpha (\Phi_P+\Phi_D) - 2\epsilon_{\alpha z \beta} \Omega_0
u_{\beta} - \frac{1}{N} \d_{\beta} P_{\alpha \beta}, \\
D_t P_{i j} &= -  P_{i \gamma}\d_\gamma u_j - P_{j \gamma} \d_{\gamma}
u_i -P_{ij} \d_{\gamma} u_{\gamma} -2\Omega_0(
\epsilon_{izk}P_{kj}+\epsilon_{jzk}P_{ki}) + Q_{ij},\label{p2}
\end{align}
in a shearing sheet for a general collision term,
$Q_{ij}=Q_{ij}(N,\mathbf{P})$. 

Now suppose Eq.'s \eqref{p1}--\eqref{p2} admit a homogeneous
steady state characterised by the equilibrium number density $N_0$,
Keplerian shear $\u_0=(0,-3\Omega_0\,x/2)$ and equilibrium pressure tensor
$\mathbf{P}^0$. Let us perturb this solution with a small axisymmetric disturbance:
\begin{align*}
 N &=N_0+\hat{N}(x,t), \\
 \u&= -(3\Omega_0\,x/2)\ey +\{u(x,t)\ex +v(x,t)\ey\}, \\ 
\mathbf{P}&=\mathbf{P}^0+ \hat{\mathbf{P}}(x,t),
\end{align*}
for $|\hat{N}|\ll N_0$, $|\nabla(u,v)|\ll \Omega_0$ and $|\mathbf{\hat{P}}| \ll
|\mathbf{P}^0|$,
 then take the solution of the linearized
Poisson equation to be
$$ \hat{\Phi}_D= -\frac{2 \pi Gm}{|k|} \hat{N}, $$
in which $k$ is the wavenumber of the harmonic variation of $\hat{N}$
(Binney and Tremaine 1987).

 Next we nondimensionalise: time according to the orbital timescale,
 $t=t^*/\Omega_0$; space like $\mathbf{x}=
(C_0/\Omega_0) \mathbf{x}^*$; surface density by $N_0$; velocity by
$C_0$ and the pressure tensor by $N_0 C_0^2$. On dropping the stars
 and linearising we acquire
\begin{align*}
\d_{t} \hat{N} &= -\d_{x} u, \\
\d_{t} u      &= ( 2g/|k|) \d_{x} \hat{N}+  2v -\d_{x} \hat{P}_{xx}, \\
\d_{t} v      &=  -\tfrac{1}{2}u - \d_{x}\hat{P}_{xy}, \\
\d_{t} \hat{P}_{xx} &= -3 P_{xx}^0 \d_{x}u + 4 \hat{P}_{xy} + \hat{Q}_{xx}, \\
\d_{t} \hat{P}_{xy} &= -2 P_{xy}^0 \d_x u - P_{xx}^0 \d_x v
-\tfrac{1}{2} \hat{P}_{xx} + 2 \hat{P}_{yy} + \hat{Q}_{xy} \\
\d_{t} \hat{P}_{yy} &= -P_{yy}^0 \d_{x}u -2 P_{xy}^0 \d_x v - \hat{P}_{xy} +
\hat{Q}_{yy}, \\
\d_t \hat{P}_{zz} &= -P_{zz}^0 \d_x u + \hat{Q}_{zz}.
\end{align*}
The perturbed collision terms take the form 
\begin{align*}
 \hat{Q}_{ij} &= \left(\frac{\d Q_{ij}}{\d N}\right)_0\, \hat{N} +
  \left(\frac{ \d Q_{ij}}{\d P_{xx}}\right)_0\, \hat{P}_{xx}
+ \left(\frac{
  \d Q_{ij}}{\d P_{xy}}\right)_0\, \hat{P}_{xy}\\
& \hskip5cm +
  \left(\frac{ \d Q_{ij}}{\d P_{yy}}\right)_0\, \hat{P}_{yy} 
+ 
\left(\frac{ \d Q_{ij}}{\d P_{zz}}\right)_0\, \hat{P}_{zz},
\end{align*}
and self-gravity is expressed in the parameter $g$ defined by
\begin{equation}
g= \frac{\pi G N_0 m}{ C_0 \Omega_0},
\end{equation}
 which is
  the inverse of
  the Toomre $Q$ parameter.

 We perform a Fourier decomposition into radial modes for each
  perturbed quantity so that they are $\propto \text{exp}(ikx+ s
  t)$, where the growth rate is $s$, and $k=2\pi/\lambda$ is the radial
  wavenumber. An algebraic
eigenvalue problem proceeds for the growth rate: $
  \Sigma_{ij}\,z_j=s\,z_i$, where $z_i$ is
  the eigenvector $(\hat{N}, u, v,\hat{P}_{xx},\hat{P}_{xy},\hat{P}_{yy},\hat{P}_{zz})$ and $\Sigma_{ij}$ is the $7\times 7$ matrix
  governing the linear evolution of the system. Enforcing solvability returns
a seventh-order dispersion relation subject to $g$, the equilibrium
  values of $P_{ij}$ and the
  derivatives of $Q_{ij}$:
\begin{multline} \label{E20}
-s^7 +\textsf{A}s^6
 +(\textsf{B}k^2+\textsf{C})s^5+(\textsf{D}k^2+\textsf{E})s^4+(\textsf{F}k^4+\textsf{G}k^2+\textsf{H})s^3
 \\
+ (\textsf{I} k^4+\textsf{J} k^2+\textsf{K})s^2 + (\textsf{L} k^4+
 \textsf{M} k^2 + \textsf{N}) s + 
(\textsf{P} k^4+\textsf{Q} k^2)\\
+g|k|\left(  2s^5 + \textsf{R} s^4 +(\textsf{S} k^2+
 \textsf{T}) s^3+(\textsf{U}k^2+\textsf{V}) s^2
+(\textsf{W}k^2+\textsf{X}) s + \textsf{Y}k^2 \right)=0.
\end{multline}
The terms above are partitioned into those which occur independently of
self-gravity 
(coefficients $\textsf{A}$--$\textsf{Q}$) and those which arise if it is
included ($\textsf{R}$--$\textsf{Y}$).
The coefficients are real and do not depend on $g$.

\subsubsection{Viscous instabilities on long wavelengths}
The analysis of \eqref{E20} we initially limit to wavelike instabilities on large scales,
$0<k\ll 1$, i.e. wavelengths $\gg H$. And so as to capture how the various
roots scale with wavenumber we expand $s$ in $|k|$.
For $s=\mathcal{O}(k^2)$,
\begin{equation} \label{genvisc}
s = -\frac{\textsf{Q}}{\textsf{N}}\, k^2 + \mathcal{O}(|k|^3),
\end{equation}
which we identify as the potential viscous instability. Its criterion for
instability can be determined from the sign of
$(\textsf{Q}/\textsf{N})$, which is
independent of the disk's self-gravity. 

For $s= \mathcal{O}(1)$,
Eq.\eqref{E20} yields the sixth order polynomial equation,
\begin{equation} \label{E21}
-s^6 + \textsf{A} s^5+\textsf{C}s^4+\textsf{E}s^3+\textsf{H}s^2+\textsf{K}s+\textsf{N}=0.
\end{equation}
 The quadratic $(1+s^2)$ factors \eqref{E21} on
account of the two identities,
\begin{equation} \label{split}
\textsf{K}-\textsf{E}+\textsf{A}=0, \qquad \qquad \textsf{N}-\textsf{H}+\textsf{C}+1 = 0
\end{equation}
 (which may be verified by computer algebra). So, to leading order, two modes
possess growth rates $s=\pm i$. These epicyclic oscillations we identify as the
potentially overstable modes. 

 Without loss of generality we let $s= i + p |k| + q k^2+\mathcal{O}(|k|^3)$ and return to
\eqref{E20}, equating terms of order $k$. This obtains
\begin{align*}
 p &=
 \frac{(\textsf{V}-\textsf{R})+(\textsf{T}-\textsf{X}-2)i}{(7+5\textsf{C}-3\textsf{H}-\textsf{N})+(6\textsf{A}-4\textsf{E}+2\textsf{K})i}\,g, \\
&= - i g. 
\end{align*}
At order $k^2$ we collect the contributions to $q$ from self gravity and
find them equal to 
\begin{align*}
-\frac{ (15\textsf{A}-6\textsf{E}+\textsf{K}-2\textsf{V}+4\textsf{R})+ 
(-11-10\textsf{C}+3\textsf{H}+\textsf{X}-3\textsf{T})i }{(-7+5\textsf{C}-3\textsf{H}+\textsf{N})+(6\textsf{A}-4\textsf{E}+2\textsf{K})i}\,g^2=-\frac{1}{2}\,i\,g^2
\end{align*}
Computer algebra is required to establish both these identities.
They reveal that the total effect of self-gravity on the overstable modes up to $k^2$ for $k\ll 1$ lies
  only in altering the
  wave frequency. The real part of $s$ (emerging at order $k^2$) is unaffected. The criterion for overstability,
  in fact, depends on the sign of
\begin{equation} \label{genov}   
\Delta=(\textsf{J}-\textsf{P}-\textsf{D})(7+5\textsf{C}-3\textsf{H}+\textsf{N})+(\textsf{G}-\textsf{B}-\textsf{M})(6\textsf{A}-4\textsf{E}+2\textsf{K}).
\end{equation}
If $\Delta>0$ the disk is viscously overstable.

These results mirror those of a viscous fluid disk in the
long-wavelength limit, 
as studied by Schmit and Tscharnuter 1995, Spahn
\emph{et al.} 2000, and Schmidt \emph{et al.} 2001. Similarly to a fluid
disk we expect self-gravity to extend the range of parameters for which the viscous instability
and overstability occur, when $k$ takes intermediate values. Without self gravity the modes of longest
wavelength are the most susceptible to both instabilities. But
inclusion of
self gravity can induce the instabilities on intermediate lengthscales,   
$\lambda_1<\lambda<\lambda_2<\infty$, as well, where both $\lambda_1$
and $\lambda_2$ are in the neighbourhood of the Jeans length.

\subsubsection{Thermal Modes}

In actual fact Eq.\eqref{split}
represents the decoupling of the dynamic modes from the thermal modes
at leading order when $k\ll 1$. This is because the $(1,1)$ minor of
$\Sigma_{ij}$ is
\begin{align} \label{part}
\left( \begin{matrix} 
                              \mathbf{E}  && \mathbf{0} \\
                               \mathbf{0}  && \mathbf{T} 
                        \end{matrix} \right) + \mathcal{O}(k)
\end{align}
where $E_{ij}$ is the two-by-two `epicyclic block' and
$T_{ij}$ is the four-by-four 
`thermal block':
\begin{align*}
\mathbf{E} = \left( \begin{matrix}
                      0 && 2 \\
                      -\tfrac{1}{2} && 0 
                     \end{matrix} \right), \qquad 
\mathbf{T} = \frac{ \d (Q_{xx},Q_{xy},Q_{yy},Q_{zz})}{\d
                      (P_{xx},P_{xy}, P_{yy}, P_{zz})} +  \left( \begin{matrix}
                      0 && 4 && 0 && 0 \\
                      -\tfrac{1}{2} && 0 && 2 && 0 \\
                      0 && -1 && 0 && 0 \\
                      0 && 0 && 0 && 0 
                     \end{matrix} \right).
\end{align*}
We have partitioned the latter matrix into a
                      collisional component, $C_{ij}$ (the Jacobian), and a part incorporating the
                      effects of rotation and shear, $H_{ij}$.  We presume these components
                       scale like
                       $|\mathbf{C}|/|\mathbf{H}| \sim
                      \omega_c/\Omega$, i.e. the
                      hydrodynamic limit will be dominated by
                      collisional processes, and the collisionless
                      limit by the anisotropizing effects of rotation
                      and shear. Note that, to leading order in $k$,
                      Eq.\eqref{part} and the thermal partition also hold for a dense gas
                      model in which the
                       terms for collisional production of $P_{ij}$
                       and collisional
                      flux of momentum are arbitrary functions of $N$,
                      $P_{ij}$ and the rate of strain tensor.

The 4 modes associated with the
                      thermal block $T_{ij}$ include the energy
                      mode, familiar from hydrodynamics (see
                      Schmidt \emph{et al.} 2001), and three others which emerge from
                      the extra degrees of freedom arising from the
                      anisotropy. 
                      In the collisioness limit, $|\mathbf{C}|\ll 1$,
                       we find     
                      two modes
                      with $s=\pm
                      2i+\mathcal{O}(|\mathbf{C}|)$, and two with
                      $s= \mathcal{O}(|\mathbf{C}|)$. The former pair, oscillating at twice
                       the epicyclic frequency, possess, to leading order, the
                      eigenvectors $(0,0,0,4,2i,-1,0)$ and
                      $(0,0,0,-4,2i,1,0)$. These
                           correspond to a perturbation of
                      the velocity ellipsoid which is comprised of
                       the oscillation of the
                      horizontal principal axes perturbations, and their rotation
                      with respect to the equilibrium orientation
                      angle, $\delta$.  
 One of the $s= \mathcal{O}(|\mathbf{C}|)$ modes
                      we presume coincides with the energy mode
                      sketched at the end of Section 2.3. The other is
                      a mode affiliated with the relaxation
                      of anisotropy,
                      the features of which cannot be determined until
                      we specify
                      $Q_{ij}$.

In the hydrodynamic limit the thermal modes, to leading order, depend solely on the
                      collisional dynamics, their growth rates
                      coinciding with the
                      eigenvalues of $C_{ij}$. These cannot be
                      ascertained until $Q_{ij}$ is known;
                       but if it is presumed that collisions work
                      to destroy anisotropy then we might expect
                      the four modes in this limit to correspond to the
                      hydrodynamic energy mode plus three 
                      `relaxation' modes which express the decay of the
                      deviatoric parts of the stress. This assumption
                      may then be employed to determine a plausible approximation to
                      the collision term, which, in fact, reproduces the
                      generic Krook model proposed by Shu \emph{et al.} 1985
                      .  
                      This is clearer if
                      we solve for $C^2$ and $\Pi_{ij}$, rather than
                      the pressure tensor, $P_{ij}$, in which case we
                      write $\widetilde{\Sigma}_{ij}\widetilde{z}_{j}=
                      s\widetilde{z}_i$, where
                      $\mathbf{z}=(\hat{N},u,v,\hat{C}^2,\hat{\Pi}_{xx},\hat{\Pi}_{xy},\hat{\Pi}_{yy})$.                   
                      First we adopt a simple model where
                       the anisotropic parts of the stress
                      decay independently of each other at the rate
                      $\zeta_\text{A}$ when in the
                      collisional limit, but where the
                      thermal energy mode decays at the rate
                      $\zeta_\text{T}$. 
                      Then the revised collision block,
                      $\widetilde{C}_{ij}$ will take the simple form: 
                     $\widetilde{\mathbf{C}}=\text{diag}(-\zeta_\text{T},-\zeta_\text{A},-\zeta_\text{A},-\zeta_\text{A})$, which
 suggests the formula
$$ Q_{ij} = \tfrac{2}{3}\,F(N,C) N C^2\,\delta_{ij} -
G_{ij}(N, \Pi_{ij}) , $$
where $(\d_{C^2}F+F)_0=\zeta_\text{T}$, $G_{kk}=0$ and
$G^0_{ij}=-\zeta_\text{A}\Pi_{ij}^0$. If $\Pi_{ij}$ is small we may expand
$G_{ij}$ in $\Pi_{ij}$ which reproduces a variant of SS85
 (for which $\zeta_\text{A}=\omega_c$ and
$F=\omega_c(2+\varepsilon^2)/3$).
But as is argued in Shu \emph{et al.} 1985 the magnitude of the relative
motions of the particles should also impact on the rate of
thermalisation; for example, gentle or glancing collisions randomise particle
velocities less efficiently than violent head-on collisions. The
importance of this detail may
be approximately quantified by the magnitude of $C$ (mediated by $\varepsilon$
if preferred), which leads us to connect the rate of anisotropic relaxation to the velocity
dispersion via the first column of
$\widetilde{C}_{ij}$. The collisional block will hence be no longer diagonal, and
then $G_{ij}$ will be also a function of $C$. The formula which ensues
for $Q_{ij}$ is the generalisation of
Shu \emph{et al.}'s `generic
Krook model', with which it agrees to first order in anisotropy. 

The vertical structure of the disk may introduce anisotropy into
 the collisional dynamics once vertical averaging is accomplished. In
 this case $F$ and $G_{ij}$ may depend on $\Pi_{zz}$ as well.

\subsection{SS85 Model Stability Analysis}
We perturb about the steady state presented in Eq.'s
\eqref{eqm}--\eqref{E13}, and then linearise and non-dimensionalise
analogously to the previous section but with
$\hat{C}=C_0\,\hat{C}^{*}$ and $\omega_c=\Omega_0\,\omega_c^*$. 
The stars will be henceforth dropped. 

 To compute the
contribution from the perturbed collision term we expand both $\varepsilon^2$
and $\omega_{c}$ in Taylor series about $C^2_{0}$ and $N_{0}$
respectively. In doing so we have neglected any possible variation in the collision
frequency due to changes in the disk thickness (refer to the
discussion of Section 2.3), an approximation which simplifies
the analysis.
This introduces the parameters: 
$e$ (cf. Eq.\eqref{e}) and 
$$\omega_{0}'\equiv
(d\,\ln\omega_{c}/d\, \ln N)_{0}.$$ 
If dense effects were properly included then
 a measure of the influence of nonlocal effects would appear in a
 third parameter, $R\equiv a\Omega/v_c$. But we
 assume it negligibly small. Moreover, there being no other velocity
 scale, all instances of $v_c$
vanish in the non-dimensionalisation.
Consequently, the perturbed, linearised and
non-dimensionalized collision term specific to the SS85 model is
$$ \hat{Q}_{ij}= \omega_0\left[
  \tfrac{1}{3}(\omega_0'+1)(\varepsilon^2-1)\,\delta_{ij}+\Pi_{ij}^0\right]\hat{N}
+\tfrac{1}{3}\omega_0(\varepsilon^2-1+\varepsilon^2\,e)\,\delta_{ij}\hat{C}^2+\omega_0\hat{\Pi}_{ij}.$$
 
As in the general case we have an eigenvalue problem for the growth
rate $s$, subject now to the three parameters: $\omega_0$ which sets
the equilibrium, and $e$ and $\omega_0'$ which measure its linear
response. The coefficients of the resulting characteristic
polynomial in \eqref{E20} are functions
of these three and are listed in Appendix A.

The stability problem is more tractable than Eq.\eqref{E20} threatens. In
fact analytic stability criteria can be derived for every
instability the system exhibits in the long (and short) wavelength
limits.
That said the dispersion relation could be simplified substantially if we put into use
Eq.\eqref{colsig} and the average of \eqref{bridge}, in which case $\omega_0'=1$ and $e\approx-0.234,\,-0.20,\,-0.19$. But we
leave these as free parameters. Firstly, doing so
provides a means by which the collision frequency's
enhancement by the disk's self-gravity can be included. This effect should
ensure $\omega_c$ is superlinear in $N$ and thus
$3/2>\omega_0'>1$. 
Secondly, leaving $\omega_0'$ and $e$ open frees the
system to exhibit the viscous overstability.

\subsubsection{The Viscous Instability}

To begin we restrict the analysis to large scales,
$0<k\ll 1$, and suppose $s = \mathcal{O}(k^2)$.
 Given $e<0$, it follows from \eqref{N} that
$\textsf{N} < 0$; 
 so, from \eqref{genvisc}, the criterion for
instability is $\textsf{Q}>0$ which, on employing \eqref{Q}, obtains
\begin{equation} \label{viscinst}      
\omega_{0 }' > \frac{-e\,(1+\omega_0^2) (11+ 2
  \omega_{0}^2)}{18\omega_{0}^2+e\,(1+\omega_0^2)(11-2\omega_0^2)},
\end{equation}
when $\omega_{0}> \widetilde{\omega}$, where $\widetilde{\omega}$ is the
nondimensionalised turning point of $\nu$, introduced in \eqref{turn}.
 This assures us that the instability is
indeed the kinetic analogue of the viscous instability: if we substitute the expression for our
      particulate disk's effective kinematic viscosity (Eq.\eqref{E14})
      into the hydrodynamic criterion for viscous instability, 
$$ \beta+1\equiv\left(\frac{1}{\nu}\frac{d(N\,\nu)}{d\,N}\right)_0 < 0, $$
we recover precisely \eqref{viscinst}. Moreover we find that the growth
rate is
\begin{equation} \label{mono}
 s= -(\textsf{Q}/\textsf{N})\,k^2 +\mathcal{O}(|k|^3)= -3 \nu_0\, (\beta+1)k^2+\mathcal{O}(|k|^3) 
\end{equation}
on using \eqref{Q} and \eqref{N} again, which is exactly the
growth rate of the viscous mode as computed by hydrodynamics.

\vskip0.3cm
Two curves of marginal viscous stability in the $(\omega_0,\omega_0')$ plane
are plotted in Fig. 2. They diverge as $\omega_{0}$ approaches
      $\widetilde{\omega}$ from above, and asymptote to $\omega_0'=1$
      as $\omega_{0}\to \infty$. A simple calculation demonstrates that for
      $e<-9/11$ a marginal curve will asymptote to 1 from above, and for
      $e>-9/11$ from below. In the hydrodynamic regime (for a dilute
      ring) we may make the identification
    $\omega_0'=-\beta+\mathcal{O}(\omega_0^{-2})$.
 Therefore the kinetic
    criterion of viscous instability recovers the familiar
    hydrodynamic criterion, $\beta<-1$.
       
 When $e\to 0$, both the curves' turning point and $\tilde{\omega}$
      drift to the origin, rendering more of
      parameter space unstable. At $e=0$ the entire quadrant is
      unstable, which coincides with the energy instability sketched
      at the end of Section 2.4. On the other hand as $e\to
      -\infty$ (the isothermal limit) we find
      $\widetilde{\omega}\to\sqrt{11/2}$, a finite
      value.

Our discussion of $\omega_c$'s form at the end of Section 2.3 would suggest
that appropriate values of $\omega_0'$ lie between $1$ and
$3/2$. Thus $e>-9/11$ will
ensure that such disks are viscously unstable for all collision
frequencies above a critical value close to $\widetilde{\omega}$. A
quick calculation shows this value to be
$(-11e/(9+11e))^{1/2}$. The data offered by experiment
 suggest $e<-9/11$, and thus indicate a \emph{dilute} disk composed of
 hard,
icy boulders falls into this category.

 As the assumption of diluteness
fails in Saturn's rings (as discussed earlier), the validity of
these stability predictions is unclear. In fact, theories and
simulations of `dense' and low-optical thickness equilibria suggest that
the viscous instability should not appear (Araki and Tremaine 1986 and
Wisdom and Tremaine 1988).
 Only in the simulations of
Lukkari 1981 and Salo 2001 has the instability been observed, though
the former used a
scaling of the elasticity law which did not account for nonlocal
effects, and the latter was for a 2D disk.
So no physically realistic 3D simulation to date has exhibited the instability. This we
presume is
 because either the parameters are not appropriate ($a\Omega/v_c$ or $\tau$ too
 big) or the
lengthscales of the computations have not been sufficiently large to
capture the unstable modes. 

A final observation: if we let $\omega_0'=1$ and solve for $e$ we produce the stability
 criterion:
 $e>-(9/11) \omega_0^2/(1+ \omega_0^2)$, which is qualitatively the
 same as that of Stewart
\emph{et al.} 1984 who solve a simplified kinetic model for only the
optical thickness and thermal energy. So it appears that reduction is
 justified here.

\begin{figure}
\begin{center}
\scalebox{.55}{\includegraphics{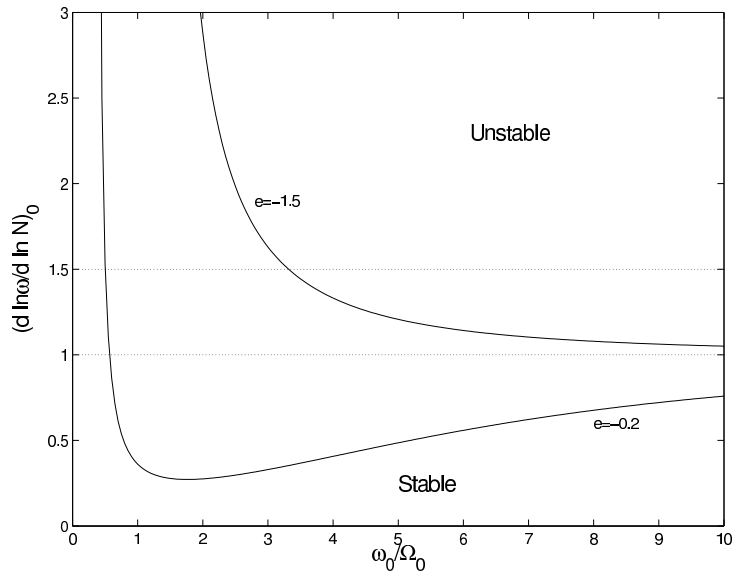}}
\caption{\footnotesize{Marginal Curves for the viscous instability in the
  $(\omega_{0},\omega_{0}')$ plane for $k\ll 1$ and different $e$ representing the
  two regimes of $e>-9/11$ and $e<-9/11$. The regions above
  the curves are unstable. The dotted lines encompass the region
  $1<\omega_0<3/2$ in which we expect to find equilibria
  characteristic of a dilute ring.}}
\end{center}
\end{figure}

\subsubsection{The Viscous Overstability}

In the long wavelength limit, $\Delta=0$ describes the curves
of marginal viscous overstability.
Substituting the
required coefficients obtains 
\begin{equation} \label{overst}
 \omega_{\text{crit}}' = \frac{ 2( g_1\,  + g_2 \,e +
  g_3\, e^2)}{9(1+\omega_{0}^2)(h_1 + h_2 \,e + h_3\,e^2)}
\end{equation}
where the $g$'s and $h$'s are the polynomials:
\begin{align*}
g_1(\omega_{0}) &= 9(-50886-75014\omega_0^2 -24440\omega_0^4-1479\omega_0^6+139\omega_0^8+10\omega_0^{10}), \\
g_2(\omega_0) &= 6(990+5855\omega_0^2+9991\omega_0^4+6014\omega_0^6+959\omega_0^8+73\omega_0^{10}+2\omega_0^{12}),\\
g_3(\omega_{0}) &=\omega_0^2(1+\omega_0^2)^2(-1851-1811\omega_0^2-100\omega_0^4+4\omega_0^6) ,\\
h_1(\omega_{0}) &= 9(-2178+833\omega_0^2+891\omega_0^4+156\omega_0^6+8\omega_0^8),\\
h_2(\omega_{0}) &= \omega_0^2(705+396\omega_0^2-351\omega_0^4-42\omega_0^6),\\
h_3(\omega_{0}) &= 2\omega_0^2(1+\omega_0^2)^2(-63+11\omega_0^2+2\omega_0^4).
\end{align*}
The modes are unstable when $\omega_{0}'>\omega_{\text{crit }}'$, and
the denominator in \eqref{overst} negative (corresponding to
$\omega_{0}\gtrsim \widetilde{\omega}$); and when
$\omega_{0}'<\omega_{\text{crit}}'$, and the denominator positive
($\omega_{0} \lesssim \widetilde{\omega}$).
Thus the stability curve comprises two branches in the
$(\omega_0,\omega_0')$ plane, as plotted
in Fig. 3. 

The right branch approaches the line
$\omega_0'=(\tfrac{2}{9} +\tfrac{2}{3}e^{-1})$ from below in the hydrodynamic
limit, $\omega_0\to \infty$, and only falls into the first
quadrant of the $(\omega_0,\omega_0')$ plane if $e<-3$. This is a
value experiments show to be too low for hard ice particles. Moreover the
maximum $\omega_0'$ for which overstability in the hydrodynamic limit is possible
is $2/9$. Thus a disk also needs to exhibit a strongly sublinear
dependence of
 $\omega_c$ on $N$ for overstability to develop, which is
 implausible. 
However, a number of formal results follow by making the identification
 $\beta=-\omega_0'$; we then obtain the hydrodynamic criterion $\beta>\beta_c=-(\tfrac{2}{9}
+\tfrac{2}{3}e^{-1})$; and if, in addition, the isothermal limit,
$e\to-\infty$, is enforced, we recover precisely the overstability criterion of Schmit and
Tscharnuter for an isothermal fluid disk without bulk viscosity:
$\beta > -2/9$.

The left branch is associated with the monotonically increasing
dependence of
equilibrium viscosity with $N$ when
$\omega_0<\widetilde{\omega}$. Fig. 3 shows these curves to
diverge when $\omega\approx \widetilde{\omega}$, and intersect the
$\omega_0=0$ axis at a finite value of $\omega_0'$, which some algebra
reveals to be $514/99-(20/297)e$. Subsequently the minimum
$\omega_0'$ for which left-branch viscous overstability is possible is
$\approx 5.192$--- a highly superlinear dependence of $\omega_c$ on $N$.
It follows that for all $\omega_0$ and $e$ there exists a `channel',
 $2/9<\omega_0'<5$, in which overstability is forbidden. The equilibria
 of a dilute ring 
 lie well within this region, hence viscous
overstability will not occur in a dilute ring.

\begin{figure}
\begin{center}
\scalebox{.55}{\includegraphics{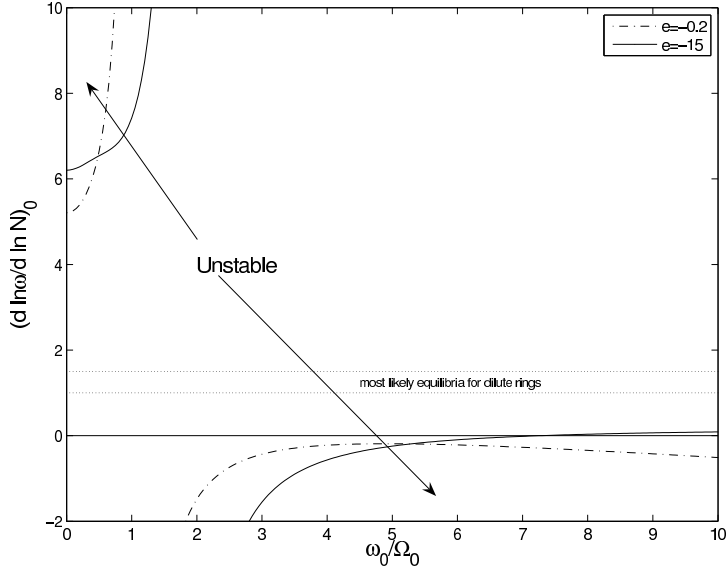}}
\caption{\footnotesize{The two branches of the marginal curves of viscous overstability in the
  $(\omega_{0},\omega_{0}')$ plane for different $e$ and with $k\ll 1$. The
  left branch intersects the $\omega_0=0$ axis at
  $\omega_0'=514/99-(20/297)e$, and the right branch asymptotes to
  the line $\omega_0'=(\tfrac{2}{9} +\tfrac{2}{3}e^{-1})$. Probable
  values of $\omega_0'$ lie in the dotted channel $1<\omega_0'<3/2$.}}

\end{center}
\end{figure}

\subsubsection{Thermal Modes}
We complete the $k\ll 1$ analysis by examining the four modes
 embodied in the SS85 thermal block
$\widetilde{\mathbf{T}}$, analogous to that appearing in
 Eq.\eqref{part}. We find that the cooling component of
this matrix is the expected,
 $\widetilde{\mathbf{C}}=\text{diag}(-\Gamma_{C^2},-\omega_0,-\omega_0,-\omega_0)$,
 where we have defined the collisional cooling rate to be
 $\Gamma\equiv\frac{1}{2}C^2 N\omega_c(1-\varepsilon^2)$,
and the subscript
 $C^2$ indicates partial differentiation with respect to squared
 velocity dispersion.

 In the hydrodynamic limit, $\omega_0\to\infty$, the heating
 contribution to $\widetilde{\mathbf{T}}$ is subdominant and  
the stress components, $\Pi_{xx}$, $\Pi_{xy}$ and $\Pi_{yy}$
 possess (to leading order)
a decay rate equal to $\omega_0$ (representing the `relaxation
 modes'), while
 the velocity dispersion,
$C^2$, decays at the rate $\Gamma_{C^2}$ (corresponding to the mode in which the energetic stability of the
 equilibrium is expressed). 

In the converse collisionless regime,
 $\omega_0\ll 1$, we recover
 the complex conjugate pair we met earlier. Their growth rates are
$$ s= \left[ \pm 2i - \tfrac{1}{176}(100+e)\omega_0+\mathcal{O}(\omega_0^2)\right]
  + \mathcal{O}(|k|),$$
and so are stable for all $e>-100$, an upper bound which is much larger than
  the values of $e$ we deem 
feasible. These two modes are analogous to those of the viscous
  overstablity, only becoming unstable when the thermal restoring
  forces are sufficiently strong (i.e. $e$ negative and large). The
  mechanism animating them we
  presume
  works to quench perturbations of the horizontal axes of the velocity
  ellipsoid, but may be so effective that the resulting `overshoot'
  may instigate an oscillation of growing amplitude. The
  instability can therefore be thought of as an `anisotropic overstability'. 

For general $\omega_0$ there is a
  single `relaxation mode' with growth rate
  $s=-\omega_0+\mathcal{O}(|k|)$,
 composed exclusively
of $\Pi_{xx}$ and $\Pi_{yy}$ perturbations in the ratio 4 to 1. 
This mode deforms the velocity ellipsoid, either vertically
  squashing or vertically dilating it, but while maintaining
   $C$ constant. Thus it is a pure rearrangement of anisotropy. Factoring out this mode from the
  characteristic polynomial of $\widetilde{T}_{ij}$ leaves a
  cubic in $s$ with a single real solution corresponding to the
growth rate of the energy mode. This is negative if:
\begin{equation} \label{therm}
 \tfrac{1}{3} \omega_0
(11+2\omega_0^2)\left[\Gamma_{C^2}-\tfrac{9}{4}\,\left(\frac{2\omega_0}{11+2\omega_0^2}\right)\right]>0.
\end{equation}
The square bracketed term we identify as the partial derivative of the energy
balance with respect to squared velocity dispersion, $C^2$, because it is composed of
the derivative of the collisional cooling function, $\Gamma$, and the
viscous heating, $\frac{9}{4}\nu_0$. Thus thermal
stability on large scales depends solely on the response of the
energy equilibrium to small fluctuations in velocity dispersion.
When we substitute 
$\Gamma_{C^2}$ into Eq.\eqref{therm} the condition simplifies to
 $e<0$, which justifies the heuristic argument offered at the end of
Section 2.4.  

\subsubsection{Instabilities on Intermediate Lengthscale}

Now we solve for general $k$. Caution should be exercised in interpreting
 the results of this analysis,
 considering the omission of the third order moments. These will
certainly dominate short-scales but should also play some part on
intermediate lengths. We will discuss their effects in more depth in
the hydrodynamic analysis of Section 4.

Let us begin without self-gravity, i.e. let
$g=0$. We seek the
marginal stability of non-oscillatory modes, and hence let $s=0$. The curves of
marginal stability for a mode of wavenumber $k$ are determined by
$$ \textsf{P}k^2+\textsf{Q}=0,$$
which derives from Eq \eqref{E20}.
The criterion $\textsf{Q}=0$ describes the marginal stability of the
longest waves and coincides with the criterion of viscous
instability. The criterion $\textsf{P}=0$ corresponds to marginal
 stability on the shortest wavelengths, and is associated with
an energy mode. 
Because we assume $e<0$ thermal stability will hold
on long wavelengths (see the analysis in Section 3.2.3) but on shorter
scales this need not be true. Outside the long wavelength regime 
density perturbations can no longer be factored out and the energy
balance's dependence on $N$ plays a part. We shall discover in the
hydrodynamic analysis that
this energy mode attempts to keep both pressure and viscous stress
constant, and its stability is related subsequently to Field's criterion (Field 1965) . 
That said the omitted
terms representing the heat flux we presume enter at higher order
than $k^2$ and should
extinguish this
instability on sufficiently short scales.
On intermediate scales instability is guaranteed if
$\textsf{Q}/\textsf{P}<0$, which refers to a hybrid mode of viscous
and thermal effects.
Again, omission of the third order moments will probably affect its
stability.

With self-gravity added the curves of marginal stability
for a non-oscillatory mode are described by
$$ \textsf{P}\,k^2 + \textsf{Y} g\,|k| + \textsf{Q}=0.$$
Solvability for $|k|$ furnishes the criterion
 $g^2> g^2_\text{crit}\equiv 4 \textsf{Q} \textsf{P}/\textsf{Y}^2$, or equivalently, $Q<Q_\text{crit}$,
where $Q$ is the Toomre parameter. The critical value $g_\text{crit}$
is a complicated function of $\omega_0$, $\omega_0'$ and $e$, which we
list in \eqref{gcrit}. 

In Fig. 4 curves of marginal stability are plotted in
$(\omega_0,\omega_0')$ space for $g=1$ and $e=-0.2$. Three
curves are plotted. Equilibria in the region above the
dotted
graph will develop \emph{viscous instability} on all wavelengths above a
critical lengthscale. This is precisely the same curve we encountered in the
long wavelength analysis, as the most unstable wavelengths are the
longest. 
The solid graph represents marginal
\emph{viscous/gravitational instability}, above which equilibria become
unstable. The unstable region includes that of the viscous
instability, which is a special case. Modes which are
viscous/gravitationally unstable possess wavelengths in an interval
$(\lambda_1,\lambda_2)$ bracketing the Jeans
length; modes which are viscously unstable possess $\lambda_2=\infty$.    
The marginal curve of viscous/gravitational instability asymptotes to
a line $\omega_0'=1-g^2/3$ for $\omega_0$ large, a fact easily
verified from Eq.\eqref{gcrit}. It follows
 that in the hydrodynamic limit the instability criterion can be
 written as $\beta<\beta_c= -(1-g^2/3)$, which agrees with
 Schmit and Tscharnuter's criterion (Eq.(35) in Schmit and Tscharnuter
 1995) for 
an isothermal disk. This fact suggests that the thermal properties of
the ring play a negligible role on the stability criterion of the most
unstable mode in the collision-dominated, hydrodynamic limit.

The dotted-dashed curve in Fig. 4 represents the \emph{quasi-thermal
instability}, which affects all modes with
wavelengths \emph{below} a critical intermediate value. Perhaps we can then
characterise it, like the viscous instability, as a special case of
gravitational/viscous instability but in which $\lambda_1=0$.
Inclusion of a heat
flux term would undoubtedly prevent $\lambda_1$ being zero.
 So in reality scales sufficiently small
will be stable.

Suppose that we take $\omega_0'=3/2$, $e=-0.234$ and $g=1$. For $k$ small the SS85 model predicts that viscous
 instability will be present
in those parts of a dilute ring in which there are more than 3.5 collisions per orbit
(i.e. $\omega_0 \gtrsim 0.555$), a critical limit which corresponds to
an optical thickness of 0.218 if we employ
 expression
\eqref{colsig}. If, however, we include
$\omega_c$'s
self-gravity
 factor, introduced in equation \eqref{ommie}, the viscous
 instability can occur for lower optical thicknesses; for example if $g=1$
 instability occurs for $\tau>0.089$.
Hence only the most tenuous of dilute rings will be stable. 

The SS85 model also
suggests that on intermediate scales quasi-thermal or viscous/gravitational instability will occur in parts of the ring
in which there are less than about 33 collisions per orbit
($\omega_0\lesssim 5.2318$), which corresponds to a critical $\tau$ of
approximately 2 (or less if $\omega_c$'s self-gravity factor is
included). This last result is dubious, of course, because of
the omission of the third order moments. The
 prevalence of quasi-thermal and viscous/gravitational instability on
intermediate scales is a question left open
until the next section. 

We will not undertake an analogous analysis of the viscous
overstability in the intermediate range, it being algebraically
intensive and not especially illuminating. In summary,
the inclusion of self gravity narrows the `channel of stability'
in
the $(\omega_0,\omega_0')$ plane, but not sufficiently to endanger the stability of equilibria 
with appropriate values of $\omega_0'$. Instability on those regions of parameter
space rendered `newly' unstable by self-gravity disturb modes on intermediate
scales only; thus the `old' boundaries of stability circumscribe
regions in which viscous overstability operates on arbitrarily long wavelengths.     
 
It should be mentioned that in the $k\gg 1$ regime the system not only evinces
the unphysical extension of the quasi-thermal instability but also two pairs of
unstable sound waves, which we also regard as unphysical. That said these
only appear for parameter values incongruent with a dilute ring of icy
particles: the first pair of modes may only be unstable when $e<-65$,
and the second when $\omega_0'<0.98$ for $e$ and $\omega_0$
sufficiently small. Thus the
thermal instability will be the primary problem if one sought to
undertake a non-linear computation of the time-dependent equations. 
\begin{figure}
\begin{center}
\scalebox{.55}{\includegraphics{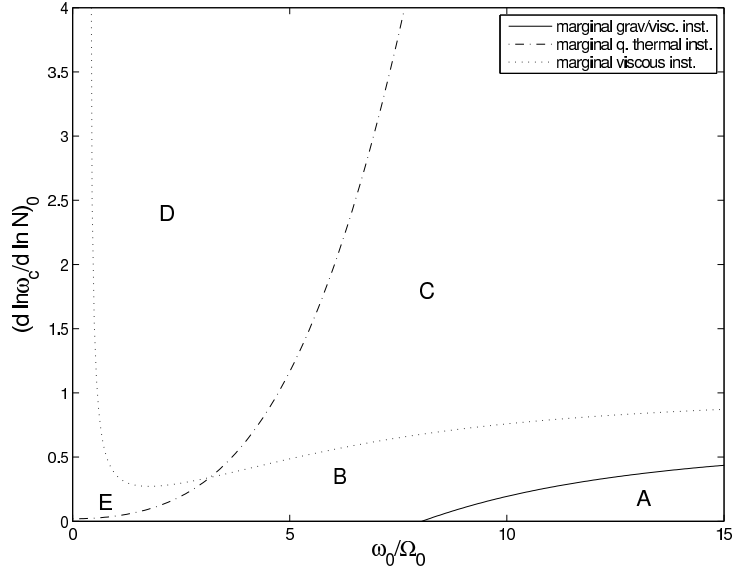}}
\caption{\footnotesize{Marginal Curves for the viscous instability,
    quasi-thermal instability and viscous/gravitational instability for
    general $k$ in the $(\omega_0$,$\omega_0')$ plane for $e=-0.2$ and
    $g=1$, as predicted by the SS85 kinetic model. Regions
    above the curves are unstable. Thus equilibria in Region A are
    stable. Region B circumscribes equilibria susceptible to
    gravitational/viscous instability on modes with $k\in (k_1,k_2)$. Equilibria in Region C feature
    viscous instability affecting modes with $k<k_3$. In Region D both quasi-thermal instability
    and viscous instability are apparent, and so every mode is unstable. And in Region E both
    quasi-thermal and gravitational/viscous instability exist on all
    wavenumbers greater than a critical value $k_4$.
    The curve
    of viscous instability asymptotes to $\omega_0'=1$ for large
    $\omega_0$, while the curve
    of viscous/gravitational instability goes to $\omega_0'=1-g^2/3$.}}

\end{center}
\end{figure}
\section{Comparison with Viscous Fluid Disk}

In this section we compare the linear analysis of our kinetic system
with that of hydrodynamics. The motivation for doing so is to
distinguish the effect of the viscous stress equation, which should
not only capture anisotropic effects but also the non-Newtonian behaviour of the stress in the range
of low and intermediate $\omega_0$. 
We produce an analysis
analogous to Schmidt \emph{et al.} 2001, and thus solve the mass conservation
equation, the Navier-Stokes equation, the temperature equation and
Poisson's equation for
a self-gravitating, viscous fluid disk in a shearing sheet. We omit
however Schmidt \emph{et al.}'s nonlocal
contributions to pressure and viscosity. So as to
best mimic our kinetic disk the transport coefficients have been set to those
computed by the SS85 kinetic equilibrium. Hence the coefficient of bulk
viscosity and heat diffusivity, $\kappa$, are
zero for the moment, the latter corresponding to the omission of the third order
moments. The shear viscosity is
\begin{equation} \label{flunu}
\nu= \frac{2 \omega_c(N)\,T}{11\Omega_0^2+2\omega_c(N)^2},
\end{equation}
where $T\equiv C^2$ is temperature.
In addition  we incorporate the `cooling function', $\Gamma$, into the
temperature equation which shall account for collisional energy losses, 
\begin{equation} \label{energy}
 \Gamma= \tfrac{1}{2}\omega_c\,N\,T\,(1-\varepsilon(T)^2).
\end{equation}
The fluid equations in the shearing sheet read:
\begin{align*}
D_t N &= -N \d_k u_k, \\
N\,D_t u_i &= -N\,\d_i (\Phi_P+\Phi_D)-\d_i P + \d_k  \Pi_{ik}, \\
\tfrac{3}{2}\,N\,D_t\, T &= -P\,\d_k\,u_k + \Psi - \Gamma,   
\end{align*}
where pressure is defined by $P\equiv NT$ and the rate of viscous heating
is $\Psi=\Pi_{kl}S_{kl}$, where $S_{ij}=e_{ij}- \d_k u_k \delta_{ij}/3$ designates the rate of deformation
tensor. The stress is Newtonian, so $\Pi_{kl}=2N\nu S_{kl}$, and $\Phi_D$ is determined
from Poisson's equation \eqref{Poisson}. It is understood these equations, and the fields
which appear in them, are vertically averaged.

The stationary state is represented by: the equilibrium surface number density
$N_0$, the Keplerian shear $\u=-\frac{3}{2}\Omega_0\,x\,\ey$, and the
equilibrium temperature $T_0$, which is computed from the energy balance.
 We
perturb, linearise and non-dimensionalise this state as earlier, expanding both $\nu$ and $\Gamma$ in Taylor series. 
On assuming wavelike perturbations $\propto e^{st+ikx}$ the system
yields the non-dimensional fourth-order dispersion relation:
\begin{align}
& s^4 + \{E_\text{T}+\tfrac{7}{3}\nu k^2\}\,s^3 \notag\\
&\hskip2cm +\{1-2g|k|+(\tfrac{5}{3}+\tfrac{10}{3}E_\text{T}
\nu_0 + 3\nu_\text{T}\,\nu_0)k^2+\tfrac{7}{3}\nu_0^2\,k^4 \}s^2 \notag\\ &
+\{E_\text{T} (1-2g|k|)+(E_\text{T}-\nu_0+ 2\nu_\text{T}\left[\tfrac{9}{2}(1+\xi)\nu_0-\tfrac{2}{3}\Gamma_\text{N}\right])k^2 \notag\\
& \hskip3.5cm -2g\nu_0 \,|k|^3+( \tfrac{7}{3}E_\text{T} \nu_0+7\nu_\text{T}\nu_0+\tfrac{5}{3})\nu_0 k^4
  \}\,s \notag \\ 
&  +\{ 2(1+\xi)\Gamma_\text{T}-2\Gamma_\text{N}\nu_\text{T}/\nu_0
-2g\,(\tfrac{2}{3}\Gamma_\text{T}+\tfrac{3}{2}\nu_\text{T})\,|k|\notag\\
&\hskip3cm +\left(\tfrac{2}{3}(\Gamma_\text{T}-\Gamma_\text{N})
-\tfrac{3}{2}(1+\xi)\nu_0+\tfrac{3}{2}\nu_\text{T}\right)k^2
\}\nu_0\,k^2=0, \label{fludisp}
\end{align}
where $\xi \equiv (\d \ln\nu/\d \ln N)_0$, $\nu_T=(\d \nu/ \d T)_0$, and
$E_\text{T} =\frac{2}{3}(\Gamma_\text{T} - \Psi_\text{T})$ which is
$2/3$ times
the partial derivative of the system's net rate of cooling with
respect to temperature. Note that $\xi$ is a partial derivative and
thus differs from $\beta$ introduced in Section 1. If heat diffusivity $\kappa$ is included other terms
materialise, the ones of relevance to our analysis being:
$$ \nu_0\,k^2(\tfrac{2}{3}\kappa k^4-\tfrac{4}{3} g\kappa |k|^3 +2(\xi+1)\kappa k^2).$$
For more details see Schmidt \emph{et al.} 2001.

In order to complete the linear theory of the fluid analogue
 we replace the four extra parameters appearing in the
dispersion relation with
\begin{align} \label{snot1} 
&\Gamma_\text{N}=\frac{9\omega_0\,(1+\omega_0')}{2(11+2\omega_0^2)}, 
&\Gamma_\text{T} =
\tfrac{1}{2}\omega_0\,\frac{9-2e(1+\omega_0^2)}{11+2\omega_0^2}, \\
&\xi=\omega_0'\,\frac{(11-2\omega_0^2)}{(11+2\omega_0^2)},
   &\nu_\text{T}=\frac{2\omega_0}{11+2\omega_0^2}. \label{snot2}
\end{align}

\subsection{Modes on Large Lengthscales}
In the long-wavelength limit, $k\ll 1$, and assuming $s=\mathcal{O}(k^2)$, the viscous
mode has growth rate:
$$ s= -2\,\frac{(1+\xi)\Gamma_\text{T}\nu_0-\Gamma_\text{N}\nu_\text{T}}{E_\text{T}}\,k^2 +\mathcal{O}(k^3) $$
 On applying
 Eq.'s \eqref{snot1}--\eqref{snot2}, curves of marginal viscous
instability are derived which are identical to those predicted by the
kinetic model in Eq.\eqref{viscinst}.
Also if we let $s=\mathcal{O}(1)$ we find the long-wavelength thermal mode:
$$ s= -E_\text{T} + \mathcal{O}(k^2).$$
which reproduces Eq.\eqref{therm}.

 The growth rate of an
overstable mode Schmidt \emph{et al.} show to be
$$ s= i -ig|k| +\tfrac{1}{2}(\gamma_i-g^2)ik^2+\tfrac{1}{2}(\gamma_r +
\tfrac{2}{3}\nu_0+3\xi \nu_0)k^2, $$
where 
\begin{align*}
\gamma_i &= -\frac{1}{1+ E_\text{T}^2} \left[
  \left(\frac{2}{3}\Gamma_\text{N}\,E_\text{T}-\frac{1+3\xi}{2}\,\nu_0\,E_\text{T}-
  \frac{2}{3} \right)\right.\\
& \left.\hskip5cm +3\nu_\text{T}\left(\frac{2}{3}\,\Gamma_\text{N}-\frac{(1+3\xi)}{2}\,\nu_0+\frac{2}{3}\,E_\text{T}\right)\right]   , \\
\gamma_r &=-\frac{1}{1+ E_\text{T}^2}\left[ \left(
  \frac{2}{3}\Gamma_\text{N}-\frac{1+3\xi}{2}\,\nu_0+
  \frac{2}{3}\,E_\text{T}
  \right)\right. \\
& \left. \hskip5cm-3\nu_\text{T}\left(\frac{2}{3}\Gamma_\text{N}\,E_\text{T}-\frac{1+3\xi}{2}\,\nu_0\,E_\text{T}-\frac{2}{3}\right)\right],
\end{align*}
in the absence of nonlocal terms.
Hence the curves of marginal overstability proceed from
$$ \xi_{\text{crit}} = \xi^* \equiv \frac{-4 \Gamma_\text{N}-4
 E_\text{T}+7\nu+4 E_\text{T}^2\nu_0+3\nu_\text{T}(4-4E_\text{T}\Gamma_\text{N}-3E_\text{T}\nu_0)}{9\nu_0(3+2E_\text{T}^2+3E_\text{T}\,\nu_\text{T})}.$$
which translates, in the parameters of our dilute ring, to
\begin{equation} \label{fluover}
\omega'_\text{crit}=\frac{(11+2\omega_0^2)(l_1+e\,l_2+e^2\,l_3)}{9(l_4+e\,l_5+e^2\,l_6)},
\end{equation}
where
\begin{align*}
&l_1 = 45(11+2\omega_0^2)^2, 
&l_2 = 6(121+183\omega_0^2+66\omega_0^4+4\omega_0^6), \\
&l_3 = 8\omega_0^2(1+\omega_0^2)^2, 
&l_4 = 9(-1331+132\omega_0^4+16\omega_0^6), \\
&l_5 = -72\omega_0^4(1+\omega_0^2),
&l_6 = 4\omega_0^2(-11+2\omega_0^2)(1+\omega_0^2)^2.
\end{align*}
These curves resemble roughly those of the kinetic model, consisting of two branches: one tightly bunched near
the collisionless limit and one which extends over the remainder of
parameter space (See Fig. 5). In the limit $\omega_0\to\infty$
the graphs asymptote, like the kinetic curves, to 
$\omega_0'= (\frac{2}{9}+\frac{2}{3}e^{-1})$. However for intermediate and lower values of $\omega_0$ the
two accounts disagree markedly.
The kinetic and fluid graphs diverge at different
values of the collision frequency, $\omega_0$. And the left branch intersects the $\omega_0=0$ axis
 at $\omega_0'\approx -0.556-0.074 e$. Hence for $e>-1.874$ overstability
 occurs for all $\omega_0'$ to the left of the singularity. This
 contradicts the kinetic
 model which predicts that no viscous overstability can occur for $2/9<\omega_0'<5.192$.   
Thus a dilute ring of icy particles
 with $e=-0.234$,
 $\omega_0'=3/2$, and $g=1$ manifests viscous overstability for all
 $\omega_0<2.2022$, or $\tau<0.541$ 

 This is an important qualitative
difference, and which we argue results from the non-Newtonian nature of the
viscous stress, in particular its nonlocality in time. This effect
 interferes with the phase synchronisation of the stress and
 density oscillations, and thus the
mechanism sustaining the overstable modes. It is less
clear how anisotropy is at work here, but we conclude it plays a smaller
effect in the linear theory.
 This effect we expect to be less appreciable for a dense gas, as the collisional contribution
to the viscous stress, which is local in time, will be of order or greater than
the free-streaming component, which is non-local in time.
 \begin{figure}
\begin{center}
\scalebox{.55}{\includegraphics{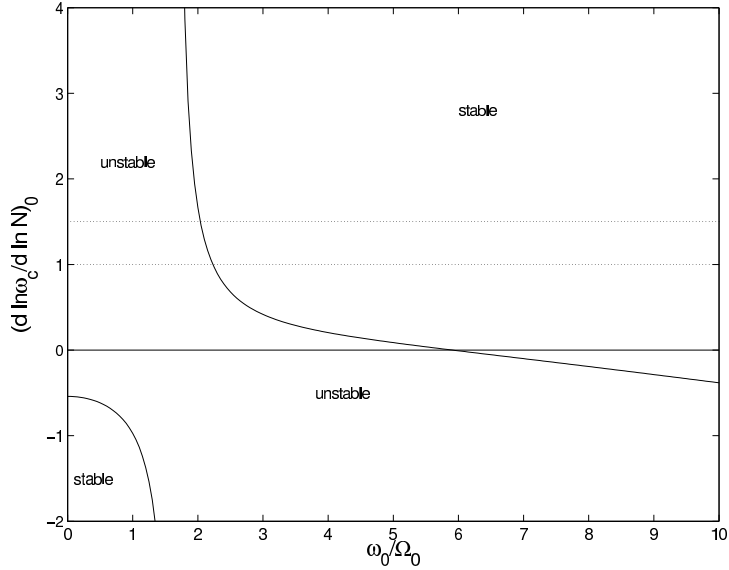}}
\caption{\footnotesize{Marginal curve of viscous overstability
    predicted by hydrodynamics for $e=-0.2$ and $k\ll 1$. The right
    branch asymptotes to the line $\omega_0'= 2/9+2/(3e)$, for
    large $\omega_0$ (in agreement with the kinetic model)
   but diverges to $+\infty$ for an intermediate
    value of $\omega_0$, as opposed to the kinetic model which diverges
    to $-\infty$. The left branch differs markedly, intersecting the
    $\omega_0=0$ axis for negative $\omega_0'$ and thus ensuring a
    large sector of parameter space is unstable. The channel
    $1<\omega_0'<3/2$ indicates the region in which we expect
    appropriate
    equilibria to fall.}}
\end{center}
\end{figure}

\subsection{General Stability Criterion for Non-Oscillatory Modes}
We derive a relation describing the marginal stability of non-oscillatory modes
from \eqref{fludisp} if we set $s=0$. It follows
that instability occurs on an interval of $k$
 if the following equation admits two solutions for $|k_c|$, 
\begin{align}
&\left[\tfrac{2}{3}(\Gamma_\text{T}-\Gamma_\text{N})
-\tfrac{3}{2}(1+\xi)\nu_0+\tfrac{3}{2}\nu_\text{T}\right]k_c^2
 -2g\,\left[\tfrac{2}{3}\Gamma_\text{T}+\tfrac{3}{2}\nu_\text{T}\right]|k_c| \notag\\
& \hskip6cm+\left[2(1+\xi)\Gamma_\text{T}-2\Gamma_\text{N}\nu_\text{T}/\nu_0\right]
=0, \label{fluin} 
\end{align}
 Eq.\eqref{fluin} is the counterpart of \eqref{mono}, the coefficients of
$k_c^2$, $|k_c|$ and $k_c^0$ corresponding to $\textsf{P}$, $\textsf{Y}$ and $\textsf{Q}$,
respectively. Thus the coefficient of
$k_c^0$ determines the onset of viscous instability and
dominates in the long wavelength limit. The coefficient of $|k_c|$
 introduces self-gravitation. The coefficient of $k_c^2$
manifests thermal effects, and is in fact equal to the
derivative of $2/3$ times the net rate of cooling,
$E=\frac{2}{3}(\Gamma-\Psi)$, with respect to temperature,
if \emph{both} pressure and the viscous stress are kept constant.
To show this we must include the variations induced by
 temperature and density displacements upon the shear rate,
$A$, which compensates for the variation in $\nu$ induced by the same
perturbations. If $\Pi_{xy}$ is to be kept constant, given a fluctuation in $N$
and $T$ at constant $P$, then
$$ \frac{\d A}{\d T} = \tfrac{3}{2}\left[
  \nu_\text{T}/\nu_0-(1+\xi)\right],$$
and therefore 
$$ \left( \frac{\d E}{\d T}\right)_{P,\,\Pi_{xy}}  = \tfrac{2}{3}(\Gamma_\text{T}-\Gamma_\text{N})
-\tfrac{3}{2}(1+\xi)\nu_0+\tfrac{3}{2}\nu_\text{T}.$$
It follows that the criterion for quasi-thermal instability on very
short scales (when Eq.\eqref{fluin} is dominated by the thermal term) is a modification of Field's
criterion $ (\d E/\d T)_P <0$ (Field 1965). But on intermediate scales, the potentially
unstable mode will be an amalgam of effects associated with the
thermal balance, self-gravitation and angular
momentum transport, all constrained to maintain constant pressure
and stress. 

The marginal curves which follow from Eq.\eqref{fluin}
match those of SS85 (described in Fig. 4); the viscous/gravitational
graph nearly coincides, and the quasi-thermal graph is qualitatively the
same. Here the discrepancies which do exist between the two models 
 we attribute to the breakdown of the assumptions of both on
 intermediate scales. Their details we omit.  

 Heat diffusivity introduces the additional terms
  $ -\frac{4}{3} g\kappa |k|^3
+\frac{2}{3}\kappa k^4$ which contribute at sufficiently small
scales. Including heat conduction will also incur a $k^2$
term, $2(\xi+1)\kappa k^2$, representing the coupling of thermal
  transport with rotation
 and viscosity. Hence the criterion of
 intermediate scale instability will be altered significantly. This
may be quantified by a
critical heat diffusivity, $\kappa_\text{crit}$
below which quasi-thermal instability can emerge on a band
  of $k$, the upper limit depending sensitively on $\kappa$. An
 expression for this quantity in the  non-self-gravitating case is presented in Eq.\eqref{kapcrit}. We find this critical
  value to be very small indeed; for instance, when
  $e=-0.234$, $\omega_0'=1$ and $\omega_0\lesssim 0.7$ then
  $\kappa_\text{crit} \lesssim 0.07$. The dilute kinetic theory
  asserts that $\nu\sim\kappa$ (Cowling and Chapman 1970), so for
  \eqref{flunu} only
  dilute rings of large optical thicknesses ($\tau \gtrsim 3$)
  should feature the quasi-thermal instability.

 When the destabilising effect of self-gravity is added, however, $\kappa_\text{crit}$ can approach feasible
values. Instability in this case we name viscous/gravitational,
though in fact the mode in question is fueled by viscous,
gravitational and thermal processes. In order to establish how
this instability impacts on the various scales we plot curves of
marginal stability in the $(\omega_0,\omega_0')$ plane for given $e$, $\kappa$
and $k$, as described by Eq.\eqref{bitch} (Fig. 6). When $\kappa
\lesssim 1$ we find that
the marginal curves fall into three categories corresponding to long,
intermediate and short scales.
For scales $k<k_{1c}$ we obtain curves
 resembling those of viscous
instability (see Fig. 2). But for scales $k_{1c}<k<k_{2c}$ there appear two
branches, the right one familiar from Fig. 4, the left arising
from non-zero $\kappa$. Instability emerges on equilibria situated
between the two. 
We conclude that heat diffusion
 confines instability on intermediate scales to a smaller area of
 parameter space than predicted otherwise; in particular, equilibria
 with low $\omega_0$ are stabilised. For
short waves, $k>k_{2c}$, the marginal curves mimic in shape
those of 
long wavelength, but as $k$ gets large $\omega_\text{crit}'\gtrsim
k^2$. Therefore if small scales are to be rendered unstable
the system needs to exhibit an inappropriately large value of
$\omega_0'$. As expected, heat diffusivity stabilizes short
scales. Expressions for $k_{1c}$ and $k_{2c}$ may be obtained by
examining the
real roots shared by the numerator and denominator of \eqref{bitch}.
     For
  $\kappa=1$, they are approximately equal to $ g\pm
  \sqrt{g^2-(6/11)\widetilde{\omega}}$, which lie on either side of
  half the Jeans length.  
 The fluid model then predicts that a
 dilute ring, with self gravity ($\omega_0'=3/2$) and $\kappa=1$,
is stable on scales $\lambda^*< 2\pi\sqrt{2/3}$, or about 5 times the
disk semi-thickness. 

It is likely a kinetic theory,
accounting for heat conductivity adequately, would
 predict similar results, though this cannot be checked until such a
 model is devised.
 However, hydrodynamics shows clearly that if
 we wish to study behaviour on
 intermediate and short scales, it is necessary to include in some way
 the third-order moments.
\begin{figure}
\begin{center}
\scalebox{.5}{\includegraphics{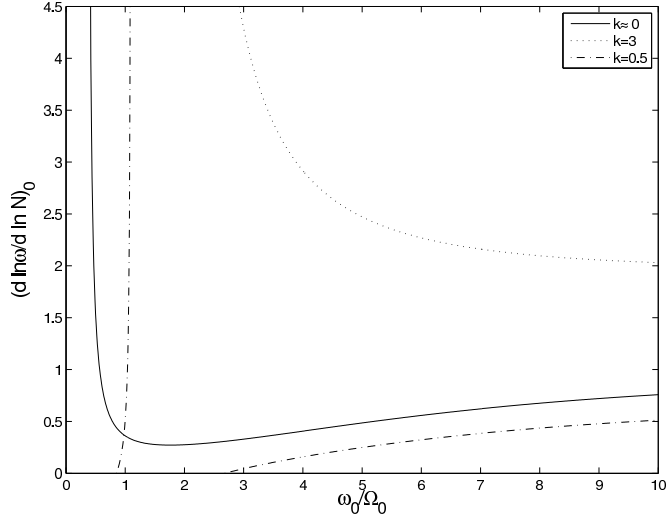}}
\caption{\footnotesize{Marginal curves of non-oscillatory instability
    for three representative values of $k$ as predicted by the
    hydrodynamical model, when $e=-0.2$, $g=1$, and
    $\kappa=1$. Equilibria
    situated
    above the $k\approx 0$ curve are unstable on the longest scales,
   but those above the $k=3$ curve are stable for scales
    $k>3$. Marginal curves for larger $k$ disappear into the top
    right. The area
    outside the region bounded between the two branches of the dotted-dashed
    curve are stable on scales $k>0.5$}}

\end{center}
\end{figure}
\section{Comparison with the Goldreich \& Tremaine Kinetic Model}

This section outlines the linear stability of the GT78 model, so as to
 compare
 the modified Krook collision term with that of
 a formal Boltzmann-type account.

 Goldreich
and Tremaine take the distribution function, $f$, to be a three-dimensional Gaussian in
velocity space with its axes corresponding to the principal axes of
the pressure tensor, $p_{ij}$. Assuming symmetry about the plane $z=0$,
the principal axis system is denoted by the unit vectors $\eee_{1}$,
$\eee_2$ and $\eee_3=\ez$, so that $\ez \sin\delta=\ex\times\eee_1$,
where $\delta$ denotes the orientation angle. The principal
values of $p_{ij}$ are accordingly defined by  $p_{kk}=n\, c_{k}^2$, where
$k=1,2,3$, and the summation convention is not applied.
The velocity distribution is then
\begin{equation}
f(\v) = \frac{n}{(2\pi)^{3/2}c_1 c_2 c_3}\,\text{exp}\left( -
\sum_{k=1}^3\frac{v_{k}^2}{2c_k^2}\right).  
\end{equation}
This form results in a much simplified collision term; see Eq (33) in
 GT78. However to facilitate better the comparison with SS85, we assume
 the value of $\varepsilon$, which appears in the
 collisional kinematics, is a quantity averaged over
 collisions, and not a constant. 
Also we note that the implicit collision frequency in the resulting $Q_{ij}$
 is $\propto \Omega \tau\,c_k/c_3 $ and thus is
dependent on the velocity dispersions, being `weighted' differently
for each direction. It does not capture self gravity or
filling factor effects in its present form-- these requiring a
superlinear dependence on $\tau$.

\subsection{Equilibrium Solution}

For the equilibrium analysis we solve the four components of the
pressure tensor equation \eqref{E8} in the principal axis frame, cf. Eqs (37) in GT78. Given $\tau$, these
determine
$\varepsilon$, $\delta$, and the axis ratios of
the velocity
ellipsoid, $c_2/c_1$ and $c_3/c_1$.

\begin{figure}
\begin{center}
\scalebox{.5}{\includegraphics{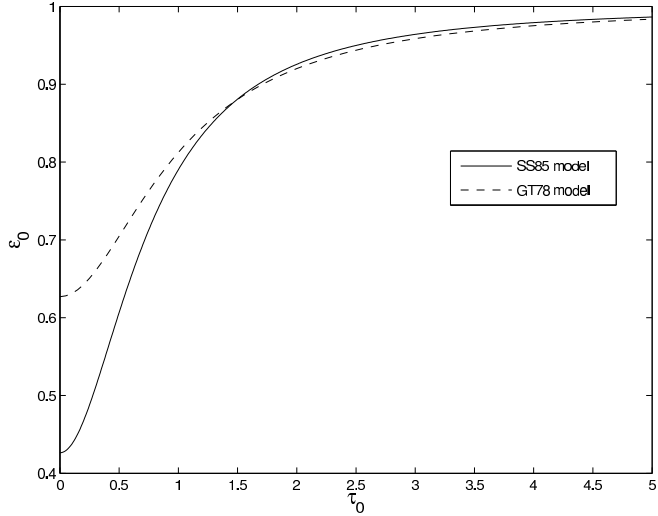}}
\caption{ \footnotesize{The energy equilibrium relations between
   the coefficient of restitution, $\varepsilon_0$, and optical thickness, $\tau_0$, as determined by the SS85 and GT78 models.}}
\end{center}
\end{figure}

The results of our numerical root finding reproduce those of GT78 and
can be observed in Fig. 7, in which we plot the $\varepsilon-\tau$ law
only.
 The SS85 relationship is plotted for comparison. Eq.\eqref{colsig}
has been used to relate $\tau$ and $\omega_c$. As mentioned earlier
the discrepancy at low $\tau$ is marked. For more discussion of this see
SS85 and Shu \emph{et al.} 1985.

\subsection{The GT78 Linear Stability}

In order to model the action of self gravity on the
collision frequency we re-parametrise the
disk equilibrium and replace optical thickness by $\omega_0$, as expressed
in \eqref{colsig}. For the purposes of the linear stability analysis
we then treat $\omega_c$ as a superlinear function of $N$. This we
admit is somewhat crude but should approximate to an acceptable degree
self-gravity's enhancement of the collision frequency, and certainly
sharpen the comparison with the SS85 model.  
  
We now perturb about the steady state calculated in the previous
section. We
then linearise and non-dimensionalise, though on account
of the complicated form of the collision term this process is rather
involved:
\begin{enumerate}
\item First, we must calculate, in
the principal axis frame, 
the three collision term perturbations, $\hat{Q}_{kk}$, in
terms of perturbations of $\tau$, $\delta$, and the principal velocity
dispersions, $c_{k}$.
\item Then the $c_k$ and $\delta$ perturbations are converted to those
 of the $\hat{P}_{ij}$ components using coordinate
transformation rules. 
\item Lastly $\hat{Q}_{ij}$ is written
in terms of $\hat{Q}_{kk}$ using the transformation
rules. 
\end{enumerate}
In this scheme 
$\varepsilon$ is assumed a function of $C^2$, where $C^2=
\tfrac{1}{3}(c_1^2+c_2^2+c_3^2)$, and is Taylor expanded about
the steady state.
After it is assumed that the perturbations are wavelike and
axisymmetric,
 a seventh-order dispersion relation emerges of the same
structure as \eqref{E20} and dependent on the three parameters,
$\omega_{0}$, $\omega_0'$ and $e$, alongside $k$.

\begin{figure}
\begin{center}
\scalebox{.5}{\includegraphics{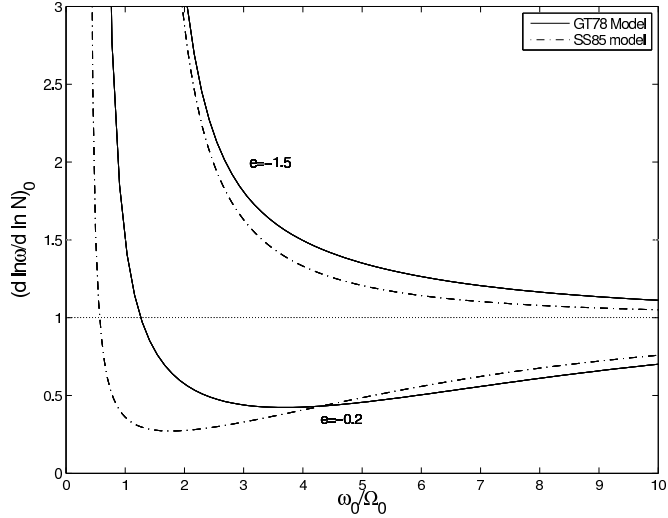}}
\caption{ \footnotesize{Comparison between the SS85 and GT78 kinetic
    models with respect to
    the marginal curves of viscous instability, for $e=-0.2$ and $e=-1.5$. Regions above the curves are unstable}}
\end{center}
\end{figure}

Given $e$, the marginal curves of the viscous instability and overstability are
numerically determined in the $(\omega_0,\omega_0')$ plane. They are plotted in Figs 8 and 9.

The curves for the viscous instability share the same qualitative
features as those of the SS85 model. Though the former
 deviate
somewhat in shape, their singularities approximately coincide,
they asymptote to the line $\omega_0'=1$, and the critical value of $e$
(which determines whether a curve approaches this line from above or below) is roughly $-9/11$.
 This suggests that not only is the viscous mode
 resistant to the effects of non-locality in time (as proven in
 the previous section) but also
 to the details of the collision term, in particular to anisotropy,
for which the Goldreich and Tremaine model accounts somewhat better. 

\begin{figure}
\begin{center}
\scalebox{.5}{\includegraphics{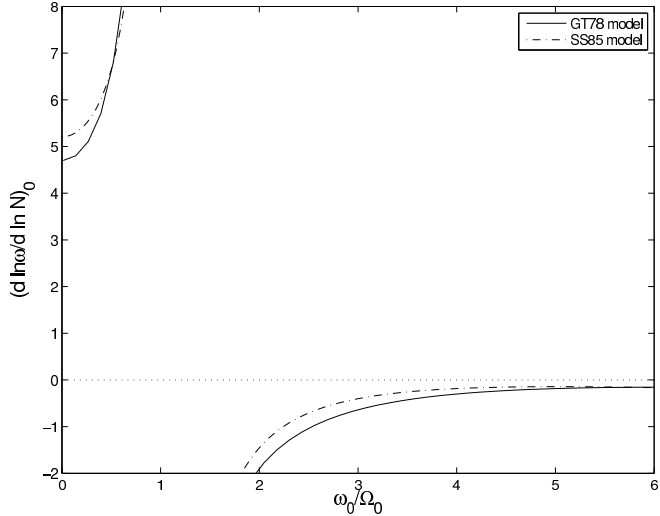}}
\caption{ \footnotesize{Comparison between the SS85 and GT78
     models with respect to the
    marginal curves of viscous overstability when $e=-0.2$. Regions
    above the left branch are unstable while regions below the right
    branch are unstable.}}
\end{center}
\end{figure}

Both branches of the curves of overstability agree in their
general features with those of SS85.
 In the large $\omega_0$ limit, the right branches asymptote to
$\omega_0'= \text{cst}$; in the isothermal limit this constant
is $2/9$, according with the SS85 and fluid models. However the
critical $e$, below which marginal curves can cross into the first
quadrant,
is closer to $-2.4$ than $-3$. Also the left branches intersect the
$\omega_0=0$ axis at slightly lower values: stability is assured in
the region left of the singularity if $\omega_0'\lesssim 4.674$, as
opposed to $\approx 5.192$.
 We conclude
that the linear stability of the viscous overstability is a little
more sensitive to the details of the collision term, and hence
anisotropy--- though not nearly as much as to
non-locality in time. As previously mentioned, the mechanism of overstability
relies on the synchronisation of two oscillations, that of the
acoustic-inertial wave, and that of a forcing which issues from the
accompanying
variation in angular momentum flux. The instability is
 vulnerable to disruption in the phase and in the
magnitude of this
forcing. A stress free to oscillate independently certainly
impacts on the former. Anisotropy may impact on the latter.

Lastly we resolve the potential `anisotropic overstable' modes for low $\omega_0$
and find them unstable for $e\lesssim -48$, far larger than predicted
by the SS85 model (though still unfeasible). This emphasises again the
fact that the two models differ most in their treatment of
anisotropy. We conclude that the less accurate SS85 model, only linear in
$\Pi_{ij}$, fails to fully model the deformations of the velocity ellipsoid.

\section{Conclusion}

In summary, we have calculated the linear stability of a patch of a
dilute planetary ring
in a shearing sheet employing a second-order kinetic theory. We have
defined `diluteness' as the insignificance of non-local effects, whose importance
we can roughly quantify by $a\Omega/c$ for lower optical depths. But throughout Saturn's rings
the velocity dispersion is low enough and particle
size large enough, for this quantity to be non-negligible, thus we
recognise that the applicability of a dilute model is restricted. Nevertheless
this work reveals the interesting effects anisotropy and
non-Newtonian stress have on the viscous instability and
overstability. It also provides a preliminary framework in
which a linear analysis of a dense gas may be undertaken.

 We
adopted a system of vertically
integrated moment equations
derived from Shu and Stewart's inelastic generalisation of the BGK
equation in their 1985 paper. The ensuing system
manifests both the viscous instability and overstability in the long-wavelength limit, and their
extensions induced by self-gravity on intervals of intermediate
wavelength. It predicts viscous or viscous/gravitational instability in dilute rings 
where particles collide about 3.5 times or more per
orbit (optical thicknesses greater than 0.2, or less depending on the importance of
self-gravity on the collision frequency). Viscous overstability does
not feature in dilute regions.

These results were compared with a hydrodynamic model incorporating the
kinetic theory's effective viscosity (evaluated at equilibrium) and cooling
function.
We find that the criterion for viscous instability is
precisely the same as that for the kinetic model, but the two criteria diverge in the
intermediate range of collision frequency for the viscous
overstability. According to the kinetics, overstability is suppressed
in this regime,
 a result in contradiction with
the fluid model which predicts the instability to be widespread. We
conclude that hydrodynamics presents a misleading
impression of the viscous overstability's prevalence on small and
intermediate ranges of $\omega_c/\Omega_0$, for a dilute ring.
This arises from the
increasingly non-Newtonian nature of the stress in this regime, an
effect the kinetic account captures but which the hydrodynamical
cannot. We stress that this incongruence will be most marked for a
dilute ring; as the collisional stress tensor is local in time, we assume the denser
the ring the better the hydrodynamic approximation.

We also undertook a linear stability analysis of the Goldreich and
Tremaine 1978 kinetic model, employing a Boltzmann-type collision term
with a triaxial Gaussian,
which should factor in anisotropic effects more thoroughly. The criteria for
viscous instability and overstability mirror qualitatively that of the
SS85 model, 
and thus suggest that the linear dynamics of a disk are
insensitive to the the precise details of the collision term. In
particular, the criterion of viscous instability matches closely that
of the SS85 model. The mechanism of this instability would thus seem relatively
robust, not only resistant to the effects of non-locality in time but
also of anisotropy. The discrepancy between the marginal curves of
viscous overstability computed by the SS85 and GT78 models is a little
more pronounced. We conclude the viscous overstability is relatively
fragile for small and intermediate $\omega_c/\Omega_0$--- very
susceptible to non-locality in time and slightly altered by anisotropy

On intermediate and short scales the SS85 model also predicted the emergence of
a quasi-thermal instability which the hydrodynamical analysis revealed
to be associated with a mode orienting the
thermal and density perturbations in such a way to keep the equation
of motion undisturbed. This mechanism thus
mimicks that operating in
Field's thermal instability (Field 1965).
We find that the instability should occur on optical thicknesses less
than 2. However hydrodynamics
shows that this instability is severely sensitive to heat conduction;
$\kappa$ need not be large to extinguish it entirely. 
Thus the omission of third-order moments in the kinetic model
bestows on it a prominence that it does not deserve. When self-gravity
is included intermediate scale instability can occur for much larger $\kappa$
 which in this case we refer to as a `viscous/gravitational
 instability'. Self-gravity,
 in addition to its important dynamic role on intermediate scales, will enhance the collision
 frequency, and render the dependence of $\omega_c$ on $N$ as steeply
 as $N^{3/2}$. In general this will make the system more susceptible
 to both the viscous instability and overstability on long lengthscales.   

Two possibly fruitful avenues which proceed from this work are: linear
stability calculations of dense and/or spinning particle disks, and
nonlinear studies of the evolution of the viscous instability and
overstability
 using the second-order
moment equations. 

In principle the techniques employed here can be
set upon kinetic models of dense and spinning rings. Detailed formulations
have been developed in Shukhman 1984, Araki and Tremaine 1986,
Araki 1988 and Araki 1991, but, unmodified, their algebraical
truculence surely prohibits all but
equilibrium calculations. 
The success of the SS85 model encourages us
that a
 simplification of the dense disk collision terms could provide a model that
 captures the essential physics, and also expedites linear and
 nonlinear analysis. 
 The results obtained with such a model would be directly pertinent to
 N-body simulations and Saturn's B-ring. It could also ascertain the validity of the fluid
 description in the dense regime.

It would be instructive also to determine the impact of the nonlinear
 evolution of the
 second-order
 moments on the development of the instabilities
discussed, as compared to the Newtonian fluid model, which so
 far has
only been explored in the isothermal case (Schmit and Tscharnuter
 1999). They may elicit interesting nonlinear behaviour from
 the viscous overstability. But an understanding of the small-scale
 instabilities, even if unphysical, is essential for the sucessful
 numerical implementation of these equations.

\textbf{Acknowledgments}
\vskip0.4cm

The manuscript was much improved thanks to the numerous and helpful comments of Juergen Schmidt. This research was funded in part by an award from the Leverhulme Trust. HNL acknowledges the financial support of the Cambridge Commonwealth Trust.

\begin{appendix}
\section{Appendix: Dispersion Relation Coefficients}
The coefficients appearing in the dispersion relation of the SS85
 model (see Eq.\eqref{E20}) are:
\begin{align}
\textsf{A} &= \tfrac{2}{3} \omega_0 \frac{e(1+\omega_0^2)-9(6+\omega_0^2)}{11+2\omega_0^2},\label{A}\\
\textsf{B} &=-4 - \frac{36}{11 + 2\omega_0^2}, \\
\textsf{C} &= \frac{
  2\,e\,\omega_0^2(1+\omega_0^2)-(55+52\omega_0^2+6\omega_0^4)}{11+2\omega_0^2},\\
\textsf{D} &= \frac{2\omega_0}{9(11+2\omega_0^2)^2}\left[
  20e(10+11\omega_0^2+\omega_0^4)+3\omega_0'(198+36\omega_0^2)\right. \notag\\
&\left. \hskip6.5cm -3(4084+1037\omega_0^2+58\omega_0^4) \right], \\
\textsf{E}&= \frac{\omega_0}{3(11+2\omega_0^2)} \left[ e(13+19\omega_0^2+6\omega_0^4)-3(80+31\omega_0^2+2\omega_0^4)\right],\\
\textsf{F}&=-\frac{12(10+\omega_0^2)^2}{(11+2\omega_0^2)^2}, \\
\textsf{G}&=\frac{2}{9(11+2\omega_0^2)}\left[
  e\,\omega_0^2\,(553+599\omega_0^2+46\omega_0^4)-27 \,e\,\omega_0'\,\omega_0^2(1+\omega_0^2)\right.
\notag\\
& \left. \hskip0.5cm +\omega_0'\,\omega_0^2\,(2349+405\omega_0^2)
  +(-3960-13785\omega_0^2-2856\omega_0^4-132\omega_0^6)  \right],\\
\textsf{H}&=
\frac{1+\omega_0^2}{3(11+2\omega_0^2)}\left[e\,\omega_0^2(17+2\omega_0^2)-(132+18\omega_0^2)\right],
\end{align}
\begin{align}
\textsf{I}&=
\frac{4\omega_0}{9(11+2\omega_0^2)^3}\left[
  2e\,(700+849\omega_0^2+156\omega_0^4+7\omega_0^6) \right. \notag \\ 
& \left. +3 \omega_0'\,(1980+558\omega_0^2+36\omega_0^4) 
  -3(21460+7344\omega_0^2+789\omega_0^4+28\omega_0^6)\right],\\
\textsf{J}&=\frac{2\omega_0}{9(11+2\omega_0^2)^2}\left[
e\,(1+\omega_0^2)(80+32\omega_0^4+\omega_0^2(478-99\omega_0')) \notag
\right.\\
&\left. 
  -3(10\omega_0^6+\omega_0^2(1977-819\omega_0')+\omega_0^4(303-135\omega_0')+4(250+99\omega_0'))\right], \\
\textsf{K}&= \frac{\omega_0}{3(11+2\omega_0^2)}\left[
  e\,(11+17\omega_0^2+6\omega_0^4)-3(44+25\omega_0^2+2\omega_0^4)\right],\\
\textsf{L}&=\frac{4\omega_0^2}{9(11+2\omega_0)^3}
\left[ e\,(1+\omega_0^2)(2240 + 20\omega_0^4
  +\omega_0^2(451-36\omega_0')-270\omega_0') \right. \notag \\
&\left. -6(5156+5\omega_0^6+\omega_0^2(1578-459\omega_0')-2034\omega_0'-9\omega_0^4(-17+2\omega_0'))\right],\\
\textsf{M}&=\frac{2\omega_0^2}{3(11+2\omega_0^2)^2}
\left[
e\,(1+\omega_0^2)(25+2\omega_0^4+\omega_0^2(45-30\omega_0')+33\omega_0')
\right.\notag \\
&
\hskip2cm\left.
  +6(\omega_0^4(-4+6\omega_0')-2(8+33\omega_0')+\omega_0^2(-47+48\omega_0'))\right], \\
\textsf{N}&=\frac{1}{3} \,e\,\omega_0^2(1+\omega_0^2),\label{N}\\
\textsf{P}&=\frac{4\omega_0^3}{3(11+2\omega_0^2)^3} 
\left[
e\,(1+\omega_0^2)(176+2\omega_0^4+\omega_0^2(43-12\omega_0')-114\omega_0')\right.
\notag \\
&\hskip8.5cm \left.+72\omega_0'(25+4\omega_0^2)\right]
,\label{P} \\
\textsf{Q}&=-\frac{2\omega_0^3}{(11+2\omega_0)^2)}
\left[e\,(1+\omega_0^2)\{2\omega_0^2(-1+\omega_0')-11(1+\omega_0')\} -18\omega_0'\,\omega_0^2\right],\label{Q}\\
\textsf{R}&= -\frac{4\omega_0}{3(11+2\omega_0^2)}\left[
e\,(1+\omega_0^2)-9(6+\omega_0^2) \right], \\
\textsf{S}&= 2+\frac{18}{11+2\omega_0^2},\\
\textsf{T}&= -\frac{4(1+\omega_0^2)}{11+2\omega_0^2}\left[ e\,\omega_0^2-22-3\omega_0^2\right],\\
\textsf{U}&= \frac{4\omega_0}{3(11+2\omega_0^2)^2}\left[
  -2e\,(10+11\omega_0^2+\omega_0^4)
+3(316+77\omega_0^2+4\omega_0^4)\right],\\
\textsf{V}&= -\frac{2\omega_0}{3(11+2\omega_0^2}\,\left[
  e\,(11+17\omega_0^2+6\omega_0^4)
-3(44+25\omega_0^2+2\omega_0^4)\right], 
\end{align}
\begin{align}
\textsf{W}&=-\frac{4\omega_0^2}{3(11+2\omega_0^2)^2}\,\left[
e\,(49+53\omega_0^2+4\omega_0^4)
-6(133+26\omega_0^2+\omega_0^4)\right] ,\\
\textsf{X}&=- \tfrac{2}{3}\,e\,\omega_0^2(1+\omega_0^2)  ,\\
\textsf{Y}&=  -\frac{4\omega_0^3}{3(11+2\omega_0^2)^2}\,\left[
  e(29+31\omega_0^2+2\omega_0^4)-18(10+\omega_0^2) \right]. \label{Y}
\end{align}

\section{Appendix: Critical Self Gravity Constant}
We present the expression for the critical value of $g$ above which
gravitational/viscous instability arises in the SS85
kinetic model. See Section 3.1.1.
\begin{equation} \label{gcrit}
g_\text{crit}= 8\,\frac{ \omega_0^3\left[h_1(\omega_0,e) + h_2(\omega_0,e)\,\omega_0' 
+ h_3(\omega_0,e)\,\omega_0'^2\right]}{(11+2\omega_0^2)^3(e(29+31\omega_0^2+2\omega_0^4)-18(10+\omega_0^2))}
\end{equation}
where
\begin{align*}
h_1 &= 
 -e^2(16+\omega_0^2)(11+13\omega_0^2+2\omega_0^4)^2, \\
h_2 &= e(11+13\omega_0^2+2\omega_0^4)
\left[e(-62-29\omega_0^2+35\omega_0^4+2\omega_0^6)-18(100+32\omega_0^2+\omega_0^4)\right],\\
h_3 &=
-6\left[216\omega_0^2(25+4\omega_0^2)-6e(-550-481\omega_0^2+91\omega_0^4+22\omega_0^6)\right.\\
  & \left.\hskip5cm +e^2(1+\omega_0^2)^2(-209+16\omega_0^2+4\omega_0^4)\right]
\end{align*} 

\section{Appendix: Critical Coefficient of Heat Diffusivity}
Here we list the critical $\kappa$ below which quasi-thermal
instability occurs on a band of wavelengths for a
viscous fluid model without self-gravity.
\begin{align} 
&\kappa_\text{crit}= \frac{ 9(1+\xi)^2
  \nu_0^2+(1+\xi)\nu_0(4\Gamma_\text{N}+4\Gamma_\text{T}-9\nu_\text{T})-8\Gamma_\text{N}\nu_\text{T}}{12(1+\xi)^2\nu_0} \notag\\
&\hskip2cm-4\frac{\sqrt{(4\Gamma_\text{N}+9(\xi+1)\nu_0)((\xi+1)\nu_0-\nu_\text{T})((\xi+1)\Gamma_\text{T}\nu_0-\Gamma_\text{N}\nu_\text{T})}}{12(1+\xi)^2\nu_0}
 \label{kapcrit}
\end{align} 

When self-gravity is present, the  marginal curves in the
$(\omega_0,\omega_0')$ plane for given $k$, $\kappa$
and $e$ may
be determined from Eq.\eqref{fluin} with heat flux
terms added. For a dilute ring with an equilibrium computed by SS85, we obtain
\begin{align} 
&\omega_c'=-\frac{11+2\omega_0^2}{3\left[
    18\omega_0^3+33\omega_0 k^2 +\omega_0
    e(11+9\omega_0^2-2\omega_0^4)+\kappa k^2(4\omega_0^4-121)\right]}\notag \\
& \times  \left[
    \omega_0(1+\omega_0^2)(3+k^3)\,e 
+2gk\omega_0(9-e(1+\omega_0^2))-\kappa k^2(11+2\omega_0^2)(3-2gk+k^2)\right].  \label{bitch} 
\end{align}

\end{appendix}

\begin{footnotesize}

\end{footnotesize}

\end{document}